\documentclass[5p,number,twocolumn]{IEEEtran}  

\usepackage{microtype}
\usepackage{graphicx}  
\usepackage[tbtags]{mathtools}
\usepackage{amssymb, amsfonts, dsfont}
\usepackage[standard,amsmath,thmmarks]{ntheorem}
\usepackage{subcaption}
\usepackage{cite}[sort,compress]
\usepackage{enumitem}
\usepackage{comment}

\usepackage{booktabs,makecell}

\newcommand*\TRANS{{\mathpalette\doTRANS\empty}}
\newcommand*\reals{\mathds{R}}
\newcommand*\X{\reals^{d_x}}
\newcommand*\Y[1]{\reals^{d_y^{#1}}}
\newcommand*\U[1]{\reals^{d_{\hat x}}}
\newcommand*\N{\mathcal N}
\newcommand*\loc{\mathrm{loc}}

\newcommand*\mmse{\mathrm{mmse}}

\newcommand*\lin{\mathrm{lin}}
\newcommand*\com{\mathrm{com}}

\newcommand*\diag{\mathrm{diag}}

\newcommand*\COLON{{\mathbin{:}}}
\newcommand*\DEFINED{\coloneqq}

\newcommand*\EXP{\mathds{E}}
\newcommand*\MATRIX[1]{\begin{bmatrix}#1\end{bmatrix}}
\newcommand*\SMATRIX[1]{\left[\begin{smallmatrix}#1\end{smallmatrix}\right]\!}

\DeclareMathOperator\VVEC{vec}
\DeclareMathOperator\COV{cov}
\DeclareMathOperator\VAR{var}
\DeclareMathOperator\ROWS{rows}
\DeclareMathOperator\Tr{Tr}

\makeatletter
\newcommand*\doTRANS[2]{\raisebox{\depth}{$\m@th#1\intercal$}}
\makeatother

\newcommand*\hide[1]{\unskip}

\theoremstyle{plain}
\theorembodyfont{}

\newtheorem{problem}{Problem}
\theoremsymbol{\ensuremath{_\square}}
\newtheorem*{example1}{Example 1}
\newtheorem*{example2}{Example 2}
\newtheorem*{example1c}{Example 1 (cont.)}
\newtheorem*{example2c}{Example 2 (cont.)}

\usepackage{xcolor}

\newcommand\DELAY{\tau}

\begin{document}

\title{Multi-agent estimation and filtering for minimizing team mean-squared
error}

\author{Mohammad~Afshari,~\IEEEmembership{Student Member,~IEEE,}
        and~Aditya~Mahajan,~\IEEEmembership{Senior Member,~IEEE}
\thanks{The authors are with the Department of Electrical and Computer
Engineering, McGill University, Montreal, QC, H3A-0E9, Canada.
Emails: {\tt\small mohammad.afshari2@mail.mcgill.ca,
aditya.mahajan@mcgill.ca}}%
\thanks{This research was supported by the Natural Science and Engineering Research
Council of Canada (NSERC).
A preliminary version of this paper was presented in the 2018 IEEE Conference
on Decision and Control (CDC)~\cite{Afshari2018b}.}}
\maketitle
                           
\begin{abstract}
  Motivated by estimation problems arising in autonomous vehicles and
  decentralized control of unmanned aerial vehicles, we consider multi-agent
  estimation and filtering problems in which multiple agents generate state
  estimates based on decentralized information and the objective is to
  minimize a coupled mean-squared error which we call \emph{team mean-square
  error}. We call the resulting estimates as minimum team mean-squared error
  (MTMSE) estimates. We show that MTMSE estimates are different from minimum
  mean-squared error (MMSE) estimates. We derive closed-form expressions for
  MTMSE estimates, which are linear function of the observations where the
  corresponding gain depends on the weight matrix that couples the estimation
  error.  We then consider a filtering problem where a linear stochastic
  process is monitored by multiple agents which can share their observations
  (with delay) over a communication graph. We derive expressions to
  recursively compute the MTMSE estimates. To illustrate the effectiveness of
  the proposed scheme we consider an example of estimating the distances
  between vehicles in a platoon and show that MTMSE estimates significantly
  outperform MMSE estimates and consensus Kalman filtering estimates.
\end{abstract}

%=============================================================================

\section{Introduction} \label{sec:intro}

Emerging applications in autonomous vehicles and decentralized control of UAVs (unmanned
aerial vehicles) give rise to estimation problems where multiple agents use
local measurements to estimate the state of the shared environment in which
they are operating and then use these estimates to act in the environment.
In the resulting decentralized estimation problems, the objective is to
minimize the weighted mean-square error between the true state and the
decentralized estimates generated by all agents. We call such a coupled
mean-square error as \emph{team mean-squared error} and the resulting
estimates as \emph{minimum team mean-squared error (MTMSE) estimates}.

For example, consider a platoon of self-driving vehicles where the estimation objective is to
ensure that the position estimates of each vehicle are close to the true position
of the vehicle and, at the same time, the difference between the position
estimates of adjacent vehicles are close to the true difference between the
positions. Or consider a fleet of UAVs (unmanned aerial vehicles) where the
estimation objective is to ensure that the position estimates of each UAV are
close to the true position of the UAV and, at the same time, the centroid of
the estimates of all UAVs is close to the true centroid of their positions. A
salient feature of these examples is that there are multiple agents who
generate state estimates based on different information and the objective is
to minimize a weighted mean-squared error between the true state and the
decentralized estimates generated by all agents. 

We first start with a simple example to illustrate that MTMSE estimates are
different from the standard MMSE (minimum mean-squared error) estimates. 
Consider a system with two agents, indexed by $i \in \{1, 2\}$, which observe the
state of nature $x \sim \mathcal{N}(0, 1)$ with noise. In particular, the
measurement $y_i \in \reals$ of agent~$i$ is 
\[
  y_i = x + v_i, \quad
  v_i \sim \mathcal{N}(0, \sigma^2),
\]
where $x$, $v_1$, and $v_2$ are independent. 

Agent~$i \in \{1, 2\}$ generates an estimate $\hat z_i = g_i(y_i) \in \reals$
based on its local measurements, where $(g_1,g_2)$ is any arbitrary estimation
strategy. The objective is to ensure that $\hat z_i$ is close to $x$ and at
the same time the average $(\hat z_1 + \hat z_2)/2$ of the estimates is close
to $x$. Thus, the estimation error $J(g_1, g_2)$ of the estimation strategy
$(g_1,g_2)$ is given by
\begin{multline} 
     \EXP[(x - \hat z_1)^2 + (x - \hat z_2)^2 ] 
    +
    \lambda \EXP\left[\left( 
      x- \frac{\hat z_1 + \hat z_2}{2}
\right)^2 \right]
\\
  = 
  \EXP\left[
    \MATRIX{ x - \hat z_1 \\ x - \hat z_2 }^\TRANS
    \MATRIX{ 1 + \frac{\lambda}{4} & \frac{\lambda}{4} \\ \frac{\lambda}{4} & 1 + \frac{\lambda}{4} }
    \MATRIX{ x - \hat z_1 \\ x - \hat z_2 }
  \right],
  % \EXP[ (Lx - \hat z)^\TRANS S (L x - \hat z) ],
  \label{eq:MMSE}
\end{multline}
where $\lambda \in \reals_{> 0}$. Naively choosing $\hat z_i$ as the MMSE estimate
of $ x$ given $y_i$, i.e., choosing 
\[
  \hat z_i = g^{\mmse}_i (y_i) 
  \DEFINED \EXP[x \mid y_i] = \frac{1}{1+\sigma^2} y_i,
\]
gives an estimation error of
\[
	J^{\mmse} = J(g^{\mmse}_1, g^{\mmse}_2) 
    = 
    2 \Bigl(\frac{ \sigma^2}{ 1 + \sigma^2 } \Bigr)
    \Bigl(1 + \frac{\lambda}{4} \cdot \frac{1 +2\sigma^2}{1 + \sigma^2}\Bigr).
    % = (OLD)
    % \frac{1}{2} (\frac{\sigma^2}{1+\sigma^2}) (5+\frac{\sigma^2}{1+\sigma^2}).
\]
%\[
%	J^{\mmse} = J(g^{\mmse}_1, g^{\mmse}_2) 
%    = 
%     2 \Bigl(\frac{ \sigma^2}{ 1 + \sigma^2 } \Bigr)
%     \Bigl(1 - \frac{\lambda}{4} \cdot \frac{3 + \sigma^2}{1 + \sigma^2}\Bigr).
%\]
This naive strategy \emph{does not} minimize the team mean-squared
error given by~\eqref{eq:MMSE}, \emph{even within the class of linear estimation
strategies}. To see this, we identify the best linear
estimation strategy. Let
\[
  \hat z_i = g^\lin_i(y_i) = F y_i
\]
where $F$ is same for both agents due to symmetry. The estimation error for 
this linear strategy is
\[
  J^\lin = J(g^\lin_1, g^\lin_2) 
  % &= 
  % 2 \bigl[ (1 - F)^2 + F^2 \sigma^2 ]
  % + \lambda \bigl[ (1-F)^2 + \tfrac12 F^2 \sigma^2 \bigr]
  % \\
  = (2 + \lambda)(1 - F)^2 + 2\Bigl(1 + \frac{\lambda}{4} \Bigr)F^2 \sigma^2
\]
%\[
%  J^\lin = J(g^\lin_1, g^\lin_2) 
%  = (2 - \lambda)(1 - F)^2 + 2\Bigl(1 - \frac{3\lambda}{4} \Bigr)F^2 \sigma^2
%\]
which is convex in $F$. The value of gain $F$ which minimizes this estimation
error is
\[
  F %= \frac{4+2\lambda}{(4 + 2\lambda) + (4 + \lambda)\sigma^2 }
  = \frac{1}{1 + \frac{1 + \lambda/4}{1 + \lambda/2} \sigma^2 }
  = \frac{1}{1 + \alpha \sigma^2},
\]
where $\alpha = (1 + \lambda/4)/(1 + \lambda/2)$. 
%\[
%  F %= \frac{4+2\lambda}{(4 + 2\lambda) + (4 + \lambda)\sigma^2 }
%  = \frac{1}{1 + \frac{1 - 3\lambda/4}{1 - \lambda/2} \sigma^2 }
%  = \frac{1}{1 + \alpha \sigma^2}
%\]
%where $\alpha = (1 - 3\lambda/4)/(1 - \lambda/2)$.
The corresponding estimation error is
\[
  J^\lin = 
  % 2 \left( \frac{\sigma^2}{1 + \alpha \sigma^2 } \right) 
  % \left( \frac{1 + \alpha^2 \sigma^2}{1 + \alpha \sigma^2} 
  %   + \frac{\lambda}{4}\cdot \frac{1 + 2\alpha^2 \sigma^2}{1 + \alpha \sigma^2}
  % \right)
  (2 + \lambda) \frac{\alpha \sigma^2}{1 + \alpha \sigma^2}.
\]
%\[
%  J^\lin = 
%  % 2 \left( \frac{\sigma^2}{1 + \alpha \sigma^2 } \right) 
%  % \left( \frac{1 + \alpha^2 \sigma^2}{1 + \alpha \sigma^2} 
%  %   + \frac{\lambda}{4}\cdot \frac{1 + 2\alpha^2 \sigma^2}{1 + \alpha \sigma^2}
%  % \right)
%  (2 - \lambda) \frac{\alpha \sigma^2}{1 + \alpha \sigma^2}.
%\]

\begin{figure}[h!bt]
  \centering
  \includegraphics[width=0.75\linewidth]{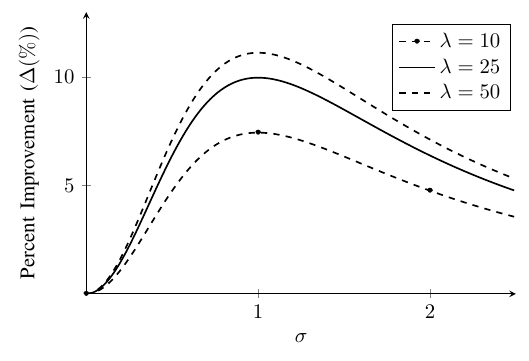}
  \caption{Comparison of the relative improvement of the best linear MTMSE estimator
    over the MMSE estimator as a function of $\sigma$ for different values of
  $\lambda$.}
  \label{fig:plot-example}
\end{figure}

Note that for large $\lambda$, $\alpha \approx 1/2$ and the relative
improvement 
\[
  \Delta \coloneqq \frac{J^\mmse - J^\lin}{J^\lin} \approx \frac{1}{2} \cdot
  \frac{\sigma^2}{(1 + \sigma^2)^2},
\]
is significant for moderate values of $\sigma$. For example, for ${\sigma = 1}$,
the relative percentage improvement is 12.5\%. A plot of the relative percentage improvement
$\Delta$ as a function of the variance $\sigma$ for different
values of~$\lambda$ is shown in Fig.~\ref{fig:plot-example}.

The relative percentage improvement 
\(
  \Delta \coloneqq (J^\mmse - J^\lin)/{J^\lin} \times 100
\)
as a function of $\sigma$ for different values of $\lambda$ is shown in
Fig.~\ref{fig:plot-example}. The improvement is significant for higher 
values 
of $\lambda$.

This significant improvement over MMSE estimates for a simple example
motivates the central question of this paper: \emph{what are the 
estimation and filtering strategies that minimize the team mean-squared
error}? We start by modeling and answering this question for estimation
in Sec.~\ref{sec:estimation}. Then, we model and answer this question
for filtering, where we assume that agents are connected over a
graph and can share their measurements over a communication graph in
Sec.~\ref{sec:filtering}. We generalize the filtering results to infinite
horizon setup in Sec.~\ref{sec:infinite}. Finally, we present examples to
illustrate that MTMSE estimates significantly outperform MMSE and consensus
Kalman filtering estimates. 

\subsection{Literature overview} \label{sec:review}
Following the seminal work of Kalman~\cite{Kalman1960}
on recursive MMSE filtering, several variations of single- and 
multi-agent MMSE filtering have been investigated in the literature.
However, as far as we are aware, there are only two references 
which have investigated estimation or filtering for the MTMSE objective~\cite{Barta1987,Andersland1996}. Both references investigated 
multi-agent filtering of a continuous time linear stochastic process. 
In~\cite{Barta1987}, each agent observes a noise corrupted measurement of 
the state and the objective is to minimize a specific form of team 
mean-squared error. The key idea of~\cite{Barta1987} is to consider an 
augmented state and observation model and formulate the team 
mean-square error as the squared norm of an appropriately defined inner 
product of these augmented variables. It is shown that team mean-squared 
filtering problem can be formulated as a Hilbert space mean-squared 
error filtering problem and, therefore, solved using an appropriate Kalman filter. The model considered in~\cite{Andersland1996} is similar 
except that each agent has multiple observation channels and, at each time,
can select which observation channel to use. The solution approach is 
similar to~\cite{Barta1987}.

Although~\cite{Barta1987,Andersland1996} are able to transform a MTMSE 
filtering problem to a Hilbert space MMSE filtering problem, the approach 
has several limitations. First, and most importantly, the approach 
of~\cite{Barta1987,Andersland1996} is only applicable to a specific 
form of MTMSE cost. The formulation of the team mean-squared error as 
a squared norm of an appropriately defined inner product does not hold for 
the more general team mean-squared error considered in this paper. 
In particular, the form of the team mean-squared error considered in 
the practical examples in Sec.~\ref{sec:examples} cannot be written 
as the squared norm of an appropriate inner product. Second, the size 
augmented state variables used in~\cite{Barta1987,Andersland1996} scales
linearly with the number of agents. In particular, for a $n$-agent MTMSE 
filtering where the state is of dimension $d_x$, the augmented state 
(and therefore the augmented estimate) is of dimension $n(d_x)^2 \times n d_x$.
Thus the resulting Kalman filter needs to keep track of $n^2 (d_x)^3 \times 
n^2 (d_x)^3$ dimensional covariance matrix. In contrast, the solution 
that we propose only requires a Kalman filter with a $d_x \times d_x$
dimensional error covariance. 
Finally,~\cite{Barta1987,Andersland1996} did not consider sharing of 
measurements among the agents. Such a sharing of measurements is a key
feature of the general filtering model that we consider in this paper.

Estimation problems with coupling between the estimates have been considered
in the economics literature~\cite{Lucas1972, Morris2002, Allen2006}. However,
in such models,  agents are strategic and want to minimize an individual
estimation objective. The solution concept is identifying estimation
strategies which are in Nash equilibrium which is different from the solution
concept of minimizing a common team estimation error considered here.

There is a rich literature on multi-agent filtering 
for distributed sensor 
fusion~\cite{Sanders1974, Sanders1978, Speyer1979, Chong1979,Hassan1978}
as well as for distributed simultaneous localization and
mapping (SLAM) in robotics~\cite{Julier2007,Li2013, Noack2017}.
There is also a rich literature on multi-agent estimation using consensus and 
gossip Kalman filters~\cite{Olfati-Saber2007,
Olfati-Saber2009,Kar2009,Cattivelli2010,Olfati-Saber2012,Battistelli2015}
(and references therein). However, all these methods only consider 
MMSE filtering. As illustrated by motivating example presented at the beginning, MTMSE estimates can be significantly different from MMSE 
estimates. So, the vast literature on multi-agent MMSE filtering 
is not directly applicable for MTMSE filtering.

\subsection{Contributions of the paper}

The salient feature of the model is that agents are informationally
decentralized and need to cooperate to minimize a common team
estimation objective. Our focus is to identify the structure of estimation
strategies that find MTMSE when the graph topology, system dynamics, and
the noise covariances are known to all agents. 

We consider the problem of minimizing the team mean-squared error in
an estimation problem where the measurements of the agents may be split into a
common measurement and local measurements.\footnote{If no such split is
possible, then the common measurement is simply empty.} Using tools from team
theory~\cite{Radner1962}, we show that the optimal MTMSE estimate is a sum of
two terms. The first term is the MMSE estimate of the state given the common
measurement. The second term is a linear function of the innovation in the
local measurement given the common measurement. Furthermore, the corresponding
gains are computed by solving a system of matrix equation, which can be
converted into a linear system of equations using vectorization. 

We then consider the problem of minimizing the sum of team
mean-squared errors over time in a filtering problem where the agents share
their measurements with their neighbors over a completely connected
communication graph. Since the graph is completely connected, the information
available at each agent can be split into common information and local
information. We show that the structure of the optimal MTMSE estimates
identified in the estimation setup continue to hold for filtering as well. We
setup an appropriate linear system with delayed observation to derive
recursive formulas for the MMSE estimate of the state based on the common
information and the innovation in the local measurements given the common
measurements. We also derive recursive formulas for computing various
covariances needed to compute the gain which multiplies the innovation term in
the optimal estimates. 

Finally, we show that under standard stabilizability and detectability
conditions, a time-homogeneous estimation strategy is optimal for minimizing
the long-term average team mean-squared error. 

%\textcolor{blue}{A remarkable feature of the team optimal decentralized
%state estimate considered in this paper is that an agent’s
%team optimal decentralized estimate is not the conditional
%mean of the state given the information at the agent; rather
%it is a linear function of the local information where the
%gain depends on the weight function of the mean square
%error. This makes team optimal decentralized estimation
%fundamentally different from centralized state estimation.
%This difference is explained in details in Sec 1.1. We argue
%in Appendix~\ref{app:one-step-delayed} that this feature partly explains the lack of
%separation between estimation and control in decentralized
%stochastic control and also explain the structure of optimal
%strategies.}

A preliminary version of this paper appeared in~\cite{Afshari2018b}, where the main result for the filtering 
problem (Theorem~\ref{thm:finite}) was stated. The proof of 
Theorem~\ref{thm:finite} relies heavily on the results for 
the estimation problem (Theorem~\ref{thm:estimation}) which 
was not included in~\cite{Afshari2018b}. Neither were the 
generalization to infinite horizon (Theorem~\ref{thm:strategy-inf}). The detailed numerical experiments
and the comparison with MMSE estimate and consensus Kalman
filtering (Section~\ref{sec:examples}), the 
detailed comparison with~\cite{Barta1987,Andersland1996} (Section~\ref{sec:intro}), the relation between the MTMSE estimates and decentralized 
control (Section~\ref{sec:disc-con}), and the trade-off between MTMSE filter complexity and 
estimation accuracy (Section~\ref{sec:disc-trade}) are new as well.  

\subsection{Notation}
Let $\delta_{ij}$ denote the Kronecker delta function (which is one if $i = j$ and
zero otherwise).
Given a matrix $A$, $A_{ij}$ denotes its $(i,j)$-th element, $A_{i\bullet}$
denotes its $i$-th row, $A_{\bullet j}$ denotes its $j$-th column, $A^\TRANS$
denotes its transpose, $\VVEC(A)$ denotes the column vector of $A$ formed by
vertically stacking the columns of $A$. Given a vector
$x$, $\|x\|^2$ denotes $x^\TRANS x$. Given matrices $A$ and $B$,
$\diag(A,B)$ denotes the matrix obtained by putting $A$ and $B$ in diagonal
blocks, and $A \otimes B$ denotes the Kronecker product of the two matrices. Given matrices $A$ and $B$ with the same number of columns,
$\ROWS(A,B)$ denotes the matrix obtained by stacking $A$ on top of $B$. Given
a squared matrix $A$, $\Tr(A)$ denotes  the sum of its diagonal elements. Given
a symmetric matrix $A$, the notation $A > 0$ and $A \ge 0$ mean that
$A$ is positive definite and semi-definite, respectively. 
$\textbf{1}_{n \times m}$ is a $n \times m$ matrix with all elements being equal to one. $\textbf{0}_n$ is a
square $n \times n$ matrix with all elements being equal to zero.
$\mathbf{I}_n$ is the $n \times n$ identity matrix. We omit the subscript 
from $\mathbf{I}_n$ when the dimension is clear from context.
We sometimes consider random vectors $X=(x_1, \dots, x_k)$ as a set with
random elements $\{x_1, \dots, x_k\}$. In particular, given two random
vectors $X = (x_1, \dots, x_k)$ and $Y=(y_1, \dots, y_m)$, we define
$X \bigcap Y$ to mean $\VVEC(\{x_1, \dots, x_k\} \bigcap \{y_1, \dots, y_m\})$.
Similarly, we use $X \setminus Y$ to mean $\VVEC(\{x_1, \dots, x_k\}
\setminus \{y_1, \dots, y_m\})$.

Given any vector valued process
$\{y(t)\}_{t \ge 1}$ and any time instances $t_1$, $t_2$ such that $t_1 \le
t_2$, $y(t_1\COLON t_2)$ is a short hand notation for $\VVEC(y(t_1), y(t_1+1),
\dots, y(t_2))$. Given matrices $\{A(i)\}_{i=1}^n$ with the same number of
rows and vectors $\{w(i)\}_{i=1}^n$,
$\ROWS(\bigodot_{i=1}^n A(i))$ and $\VVEC(\bigodot_{i=1}^n w(i))$ denote
$\ROWS(A(1), \dots, A(n))$ and $\VVEC(w(1), \dots, w(n))$, respectively.

Given random vectors $x$ and $y$, $\EXP[x]$ and $\VAR(x)$
denote the mean and variance of $x$ while $\COV(x,y)$ denotes the covariance
between $x$ and $y$.

\section{Minimum team mean-squared error (MTMSE) estimation}\label{sec:estimation}  

\subsection{Model and problem formulation} \label{sec:estimation-model}

Consider a system with $n$ agents that are indexed by the set $N =
\{1, \dots, n\}$. The agents are interested in estimating the state $x \in
\reals^{d_x}$ of nature. Agent~$i$ makes a local measurement $y_i \in
\reals^{d^i_y}$, $i \in N$. In addition, all agents observe a common
measurement, which we denote by $y_0 \in \reals^{d^0_y}$. 
We use $N_0$ to denote the set $\{0, 1, \dots, n\}$. 

The variables $(x, y_0, y_1, \dots, y_n)$ are assumed to be jointly Gaussian
zero-mean random variables. For any $i,j \in N_0$, let
\(
  \Theta_i = \COV(x, y_i)
\)
and
\(
  \Sigma_{ij} = \COV(y_i, y_j).
\)

Agent~$i \in N$ generates an estimate $\hat z_i \in \reals^{d_z^i}$
according to an estimation rule $g_i$, i.e., $\hat z_i = g_i(y_0, y_i)$. Given weight matrices $\{S_{ij}\}_{i,j \in N}$ and $\{L_{i}\}_{i \in N}$, where $S_{ij} \in \reals^{d_z^i \times d_z^j}$ and $L_i \in \reals^{d_z^i \times d_x}$, the
performance is measured by the team estimation error given by:
\begin{equation} \label{eq:cost}
  c(x, \hat z_1, \dots, \hat z_n) =
  \sum_{i \in N} \sum_{j \in N} (L_i x-\hat z_i)^\TRANS S_{ij} (L_j x-\hat z_j).
\end{equation}
Let $\hat z = \VVEC(\hat z_1, \dots, \hat z_n)$ denote the estimate 
of all agents. The team estimation error $c(x, \hat z)$ is a 
weighted quadratic function of $(L x- \hat z)$. In particular,
\begin{equation} \label{eq:cost-t-closed}
  c(x, \hat z) = (L x- \hat z)^\TRANS S (L x- \hat z),
\end{equation}
where $S$ and $L$ are given by
\begin{equation} \label{eq:cost-matrix}
S =
  \begin{bmatrix}
    S_{11} & \cdots & S_{1n} \\
    \vdots & \ddots & \vdots \\
    S_{n1} & \cdots & S_{nn}
  \end{bmatrix}
  \quad\text{and}\quad
L =
  \begin{bmatrix}
    L_{1} \\
    \vdots \\
    L_n
  \end{bmatrix}.
\end{equation}
We assume  that the matrix $S$ is positive definite.

We now present a few examples of the estimation error function of the
form~\eqref{eq:cost-t-closed}:
\begin{enumerate}
    \setlength{\parindent}{1em}
  \item Suppose $x=\VVEC(x_1,\dots,x_n)$,
where $x_i$ is the local state of agent $i \in N$. Suppose
the agents want to estimate their own local state, but at the same time, want
to make sure that the average 
$\bar z \coloneqq \frac{1}{n} \sum_{i \in N} \hat z_i$
of their estimates is close to the average
$\bar x \coloneqq \frac{1}{n} \sum_{i \in N} x_i$ of their local states.
In this case, the team mean-squared error function is
\begin{equation} \label{eq:first-cost}
  c(x,\hat z) =  \sum_{i \in N} \lVert x_i - \hat z_i \rVert^2 
  + \lambda \lVert \bar x - \bar z \rVert^2,
\end{equation}
where $\lambda \in \reals_{> 0}$. This can be written in the 
form~\eqref{eq:cost-t-closed} with
$L = \mathbf{I}$, and
\[
  S_{ij} = \bigl( \delta_{ij} + \tfrac{\lambda}{n^2} \bigr) 
  \mathbf{I}.
\]
  \item Suppose the agents are moving in a line
(e.g., a vehicular platoon) or in a closed shape (e.g., UAVs flying in a
formation) and want to estimate their local state but, at the same time,
want to ensure that the difference
$\hat d_i \coloneqq \hat z_i - \hat z_{i+1}$ between their
estimates is close to the difference
$d_i \coloneqq x_i - x_{i+1}$ of their local states.

For example when agents are moving in a line, the team
mean-squared error function is
\begin{equation} \label{eq:platoon-cost}
  c(x,\hat z) =  \sum_{i \in N} \lVert x_i - \hat z_i  \rVert^2
  + \lambda \sum_{i \in N \setminus n} \lVert d_i - \hat d_i \rVert^2,
\end{equation}
where $\lambda \in \reals_{> 0}$. This can be written in the
form~\eqref{eq:cost-t-closed} with
$L = \mathbf{I}$
 and
 \[
 S_{ij} = \begin{cases}
   (1+2\lambda)\mathbf{I}, & i=j \in \{2,\dots,n-1\} \\
   (1+\lambda)\mathbf{I}, & i = j \in \{1,n\} \\
   -\lambda \mathbf{I}, & j \in \{ i+1,i-1 \}  \\
   0, & \text{otherwise},
\end{cases}
 \]
A similar weight matrix can be obtained for the case when agents are moving in
a closed shape.

\item Suppose each agent generates an estimate $\hat z_i \in
  \reals^{d_x}$ of the state $x$ of nature and the objective is to minimize
  \[
    c(x, \hat z_1, \dots, \hat z_n) = 
    \sum_{i \in N} \sum_{j \in N} (x - \hat z_i)^\TRANS S_{ij} (x - \hat z_j).
  \]
  This can be written in the form~\eqref{eq:cost-t-closed} with $L =
  \mathbf{1}_{n \times 1} \otimes \mathbf{I}_{d_x \times d_x}$. This cost
  function is equivalent to the team mean-squared error considered
in~\cite{Barta1987, Andersland1996}.
\end{enumerate}

We are interested in the following optimization problem.
\begin{problem}\label{prob:estimation}
  Given the covariance matrices $\{\Theta_i\}_{i \in N_0}$ and
  $\{\Sigma_{ij}\}_{i,j\in N_0}$ and weight matrices $L$ and $S$, choose the
  estimation strategy $g = (g_1, \dots, g_n)$ to minimize the expected
  team estimation error $J(g)$ given by
  \begin{equation} \label{eq:cost-expected}
    J(g) \DEFINED \EXP[c(x,\hat z)].
  \end{equation}
\end{problem}

\begin{remark}
  In Problem~\ref{prob:estimation}, the system model is common knowledge among
  all agents. Thus, it may be viewed as a problem of ``centralized planning
  and decentralized execution.'' The key conceptual difficulty in the problem
  is that the estimates are generated using different information (recall that
  the information available at agent~$i$ is $(y_0, y_i)$) with the objective
  of minimizing a common coupled team estimation error given
  by~\eqref{eq:cost-t-closed}. This feature makes the
  Problem~\ref{prob:estimation} conceptually different from the standard
  estimation problem of minimizing the MMSE error. 
\end{remark}

\subsection{Optimal team estimation strategy} \label{sec:opt-est}

We define three auxiliary variables: 
\begin{itemize}
  \item All agents' \emph{common estimate of state} $x$ given the common
    measurement $y_0$ at all agents. We denote this estimate by $\hat x_0$ and
    it is equal to $\EXP[x | y_0]$. 
  \item All agents' \emph{common estimate of agent~$i$'s measurement} $y_i$ given the
    common measurement $y_0$. We denote this estimate by $\hat y_i$ and it is
    equal to $\EXP[ y_i | y_0]$. 
  \item The \emph{innovation in the local measurement of agent}~$i$ with
    respect to the common measurement. We denote this innovation $\tilde y_i$
    and it is equal to $y_i - \hat y_i$.
\end{itemize}
Let $\hat \Theta_i$ denote the covariance $\COV(x, \tilde y_i)$ and $\hat
\Sigma_{ij}$ denote the covariance $\COV(\tilde y_i, \tilde y_j)$.
From elementary properties of Gaussian random variables, we have the
following: 
\begin{lemma} \label{lem:P}
  The covariance matrices defined above are given by
  \begin{enumerate}
    \item $\hat \Theta_i = \Theta_i - \Theta_0 \Sigma_{00}^{-1} \Sigma_{0i}$.
    \item $\hat \Sigma_{ij} = \Sigma_{ij} - \Sigma_{i0} \Sigma_{00}^{-1}
      \Sigma_{0j}$.
  \end{enumerate}
  Therefore, the auxiliary variables defined above are given by
  \begin{enumerate}[resume]
    \item $\hat x_0 = \Theta_0 \Sigma_{00}^{-1} y_0$.
    \item $\hat y_i = \Sigma_{ij}\Sigma_{00}^{-1}y_0$.
    % \item $\tilde y_i = y_i - \Sigma_{ij}\Sigma_{00}^{-1}y_0$.
  \end{enumerate}
  Furthermore, we have
  \begin{enumerate}[resume]
    \item $\EXP[ x_i | y_0, y_i] = \hat x_0 + \hat \Theta_i \hat
      \Sigma_{ii}^{-1} \tilde y_i$.
      \vskip 4pt
%    \item $\hat x_i - \hat x_0$ is orthogonal to $y_0$.
%      \vskip 4pt
    \item $\EXP[ \tilde y_j  \,|\, y_0, y_i] =
      \hat \Sigma_{ji}
      \hat \Sigma_{ii}^{-1} \tilde y_i$.
  \end{enumerate}
\end{lemma}

The result follows from elementary properties of Gaussian random
variables. Then, we have the following.

\begin{theorem}\label{thm:estimation}
  The estimation strategy that minimizes the team mean-squared error
  in Problem~\ref{prob:estimation} is a linear function of the measurements.
  Specifically, the MTMSE estimate may be written as
  \begin{equation} \label{eq:opt-common-static}
    \hat z_i = L_i \hat x_0 + F_i \tilde y_i,
    \quad \forall i \in N,
  \end{equation}
  where the gains $\{F_i\}_{i \in N}$ satisfy the following system of matrix
  equations:
  \begin{equation}\label{eq:matrix-eq-static}
    \sum_{j \in N} \Big[
      S_{ij} F_j \hat \Sigma_{ji} - S_{ij} L_j \hat \Theta_{i}
    \Big] = 0,
    \quad \forall i \in N.
  \end{equation}
  If $\hat \Sigma_{ii} > 0$ for all $i \in N$, then~\eqref{eq:matrix-eq-static} has
  a unique solution which can be written as
  \begin{equation}\label{eq:F-static}
    F = \Gamma^{-1} \eta,
  \end{equation}
  \begin{align*} \label{eq:gains-main-static}
    \text{where} \quad
    F &= \VVEC(F_1, \dots, F_n), \\
    \eta &= \VVEC(S_{1 \bullet} L\hat \Theta_{1}, \dots, S_{n \bullet} L \hat \Theta_{n}),
    \\
    %S_i &= [S_{i1}, \dots, S_{in}], \\
    \Gamma &= [\Gamma_{ij}]_{i,j \in N},
    \quad\text{where }
    \Gamma_{ij} = \hat \Sigma_{ij} \otimes S_{ij}.
  \end{align*}
  Furthermore, the minimum team mean-squared error is given by
  \begin{equation} \label{eq:optimal-total-cost-static}
    J^* = \Tr(L^\TRANS S L P_0 ) -\eta^\TRANS \Gamma^{-1} \eta
    ,
  \end{equation}
  where
  \(
    S_i = [S_{i1}, \dots, S_{in}]
  \)
  and
  \(
    P_0 = \VAR(x- \hat x_0).
  \)
\end{theorem}
The proof of Theorem~\ref{thm:estimation} is presented in
Appendix~\ref{sec:one-step-proof}. 

To illustrate this result, consider the two agent example presented in the
introduction. In that model, there is no common measurement. So $\hat x_0 =
0$, $\hat y_i = 0$, and therefore $\tilde y_i = y_i$. Moreover, $\hat
\Sigma_{ij} = 1 + \sigma^2 \delta_{ij}$  and $\hat \Theta_i = 1$. Therefore,
\begin{align*}
  \Gamma_{ij} &= S_{ij}\hat \Sigma_{ij} = (\delta_{ij} + \tfrac{\lambda}{4})(1
  + \delta_{ij} \sigma^2),
  \\
  \eta_i &= S_{i1} + S_{i2} = 1 + \tfrac{\lambda}{2}.
\end{align*}
Thus, the optimal gains are
\[
  F = \Gamma^{-1} \eta = \frac{1}{1 + \alpha
  \sigma^2} \MATRIX{1 \\ 1},
\]
where $\alpha = (1 + \lambda/4)/(1 + \lambda/2)$ and the minimum team
mean-squared error is
\[
  J^* = \Bigl( \sum_{i,j} S_{ij} \Bigr) - \eta^\TRANS F = 
  (2 + \lambda) \frac{\alpha \sigma^2}{1 + \alpha \sigma^2}.
\]
Thus, we recover the results obtained by brute force calculations in the
introduction.

\begin{remark} \label{rem:structure}
  In~\eqref{eq:opt-common-static}, the first term of the estimate is the
  MMSE estimate of the current state given the common measurements. The second
  term may be viewed as a ``correction'' which depends on the innovation
  in the local measurement. A salient feature
  of the result is that the gains $\{ F_i\}_{i \in N}$ depend on the weight
  matrix~$S$.
\end{remark}

\begin{remark}
When $S$ is block diagonal, there is no cost coupling among the agents
and Problem~\ref{prob:estimation} reduces to $n$ separate problems. Thus,
the MMSE estimates $L_i \hat x_i$ are also the MTMSE estimates. 
\end{remark}

\section{Minimum team mean-squared error (MTMSE) filtering}\label{sec:filtering}  

In this section, we consider the problem of filtering to minimize
team mean-squared error when agents share information over a
communication graph. We start with a quick overview of graph theoretic
terminology.

\subsection{Overview of graph theoretic terminology}

A directed weighted graph $\mathcal G$ is an ordered set $(N,E,\DELAY)$ where $N$ is the set of nodes
and $E \subset N \times N$ is the set of ordered edges, and $\DELAY \colon E \to
\reals^k$ is a weight function. An edge $(i,j)$ in $E$ is
considered directed from $i$ to $j$; $i$ is the \emph{in-neighbor} of $j$; $j$
is the \emph{out-neighbor} of $i$; and $i$ and $j$ are neighbors. The set of
in-neighbors of $i$, called the \emph{in-neighborhood} of $i$, is denoted by
$N^-_i$; the set of out-neighbors of $i$, called the \emph{out-neighborhood},
is denoted by $N^+_i$.

In a directed graph, a \emph{directed path} $(v_1,v_2,\dots, v_k)$ is a weighted
sequence
of distinct nodes such that $(v_i, v_{i+1}) \in E$. The \emph{length} of a
path is the weighted number of edges in the path. The \emph{geodesic distance} between
two nodes $i$ and $j$, denoted by $\ell_{ij}$, is the shortest weight length of
all paths connecting the two nodes. The weighted \emph{diameter} of the graph is the largest
weighted geodesic distance between any two nodes.
A directed graph is called \emph{strongly
connected} if for every pair of nodes $i, j \in N$, there is a directed path
from $i$ to $j$ and from $j$ to $i$. A directed graph is called \emph{complete}
if for every pair of nodes $i, j \in N$, there is a directed edge
from $i$ to $j$ and from $j$ to $i$.

\subsection{Model and problem formulation}

\subsubsection{Observation Model}

Consider a \hide{discrete-time} linear stochastic process $\{x(t)\}_{t \ge 1}$, $x(t) \in
\X$, where $x(1) \sim \N(0, \Sigma_x)$ and for $t \ge 1$,
\begin{equation} \label{eq:model}
  x(t+1) = A x(t) + w(t),
\end{equation}
where $A$ is a $d_x \times d_x$ matrix and $w(t) \in \X$, $w(t)~\sim~\N(0,Q)$,
is the process noise.
There are $n$ agents, indexed by $N = \{1, \dots,
n\}$, which observe the process with
noise. At time~$t$, the measurement $y_i(t) \in
\Y{i}$ of agent~$i \in N$ is given by
\begin{equation} \label{eq:measurement}
  y_i(t) = C_i x(t) + v_i(t),
\end{equation}
where $C_i$ is a $d_y^i \times d_x$ matrix and $v_i(t) \in \Y{i}$, 
$v_i(t)~\sim~\N(0,R_i)$, is the
measurement noise. Eq.~\eqref{eq:measurement}
may be written in vector form as
\[
  y(t) = C x(t) + v(t),
\]
where $C = \ROWS(C_1, \dots, C_n)$, $y(t) = \VVEC(y_1(t), \dots, y_n(t))$, and
$v(t) = \VVEC(v_1(t), \dots, v_n(t))$.

The agents are connected over a communication graph~$\mathcal{G}$,
which is a \emph{strongly connected} weighted directed graph with vertex
set~$N$. For every edge $(i,j)$, the associated weight~$\DELAY_{ij}$ is a positive
integer that denotes the communication delay from node~$i$ to node~$j$.

Let $I_i(t)$ denote the information available to agent~$i$ at time~$t$. We
assume that agent~$i$ knows the history of all its measurements and $\DELAY_{ji}$
step delayed information of its in-neighbor $j$, $j \in N^-_i$,
i.e.,
\begin{equation} \label{info-struct}
  I_i(t) =  \{ y_i(1\COLON t) \} \odot \Big( \bigodot_{j \in N^-_i}
  \{ I_j(t-\DELAY_{ji}) \} \Big).
\end{equation}
In~\eqref{info-struct}, we implicitly assume that $I_i(t) = \emptyset$ for any
$t \le 0$.

Let $\zeta_i(t) = I_i(t) \setminus I_i(t-1)$ denote the new
information that becomes available to agent~$i$ at time~$t$. Then, $\zeta_i(1)
= y_i(1)$ and for $t>1$,
\begin{equation*}
  I_i(t) = \VVEC(y_i(t),\{ \zeta_j(t-\DELAY_{ji})\}_{j \in N^{-}_i}).
\end{equation*}
It is assumed that at each time $t$, agent $j\in N$, 
communicates $\zeta_j(t)$ to all its out-neighbors. This information 
reaches the out-neighbor $i$ of agent $j$ at time $t+ \DELAY_{ji}$.

Some examples of the communication graph are as follows.
\begin{example1}
  Consider a complete graph with $\DELAY$-step delay along each edge. The resulting
    information structure is
    \[
      I_i(t) = \{ y(1{:} t-\DELAY), y_i(t-\DELAY+1{:} t) \},
    \]
    which is the \emph{$\DELAY$-step delayed sharing information
    structure}~\cite{Witsenhausen1971}.
\end{example1}
\begin{example2}
Consider a strongly connected graph with unit delay along each edge. Let
    \(
      \DELAY^* = \max_{i, j \in N} \ell_{ij},
	\)
    denote the weighted diameter of the graph and
	\(
	  N^{k}_i = \{ j \in N : \ell_{ji} = k \}
	\)
	denote the $k$-hop in-neighbors of $i$ with $N^0_i = \{i\}$. The
    resulting information structure is
	\[
	  I_i(t) = \bigcup_{k = 0}^{\DELAY^*} \bigcup_{j \in N^k_i} \{ y_j(1\COLON t-k) \},
	\]
    which we call the \emph{neighborhood sharing information structure}.
\end{example2}

At time~$t$ agent~$i \in N$ generates an estimate $\hat z_i(t) \in \reals^{d^i_z}$ 
of $L_i x(t)$ (where $L_i$ is a $\reals^{d_z^i \times d_x}$ matrix) according to
\begin{equation*}
  \hat z_i(t) = g_{i,t}(I_i(t)),
\end{equation*}
where $g_{i,t}$ is a measurable function called the \emph{estimation rule} 
at time~$t$. The collection $g_i
\DEFINED (g_{i,1}, g_{i,2}, \dots)$ is called 
the \emph{estimation strategy} of
agent~$i$ and $g \DEFINED (g_1, \dots, g_n)$ is 
the \emph{team estimation strategy profile} of all agents.

\subsubsection{Estimation Cost}

Let $\hat z(t) = \VVEC(\hat z_1(t), \dots, \hat z_n(t))$ denote the estimate 
of all agents. As in Sec.~\ref{sec:estimation}, we assume that the estimation
error $c(x(t), \hat z(t))$ is a 
weighted quadratic function of $(L x(t)- \hat z(t))$ of the form
\begin{equation} %\label{eq:cost-t-closed}
  c(x(t), \hat z(t)) = (L x(t)- \hat z(t))^\TRANS S (L x(t)- \hat z(t)).
\end{equation}
Examples of such estimation error functions were given in
Sec.~\ref{sec:estimation-model}.

\subsubsection{Problem Formulation}

It is assumed that the system satisfies the following assumptions.
\begin{enumerate}
  \item[\textbf{(A1)}] The cost matrix $S$ is positive definite.
  \item[\textbf{(A2)}] The noise covariance matrices $\{ R_i\}_{i \in N}$ 
  are positive definite and $Q$ and $\Sigma_x$ are positive semi-definite.
  \item[\textbf{(A3)}] The primitive random variables $(x(1), \allowbreak \{w(t)\}_{t \ge
      1}, \allowbreak \{v_1(t)\}_{t \ge 1}, \allowbreak \dots, \{v_n(t)\}_{t \ge
    1})$ are independent.
  \item[\textbf{(A4)}] For any square root $D$ of matrix $Q$ such 
  that $D D = Q$, $(A,D)$ is stabilizable.
  \item[\textbf{(A5)}] $(A,C)$ is detectable.
\end{enumerate}

We are interested in the following optimization problem.
\begin{problem}[Finite Horizon] \label{prob:finite}
  Given matrices $A$, $\{ C_i \}_{i \in N}$, $\Sigma_x$, $Q$, $\{ R_i
  \}_{i \in N}$, $L$, $S$, a \emph{communication graph} $\mathcal{G}$ (and
  the corresponding weights $\DELAY_{ij}$), and a horizon $T$, choose a
  team estimation strategy profile $g$ to minimize $J_T(g)$ given by
 \begin{equation} \label{eq:total-cost-finite}
   J_T(g) = \EXP^{g} \bigg[ \sum_{t=1}^{T} c(x(t), \hat z(t)) \bigg].
 \end{equation}
\end{problem}

\begin{problem} [Infinite Horizon]\label{prob:infinite}
  Given matrices $A$, $\{ C_i \}_{i \in N}$, $\Sigma_x$, $Q$, $\{ R_i
  \}_{i \in N}$, and a \emph{communication graph} $\mathcal{G}$ (and the
  corresponding weights $\DELAY_{ij}$), choose a team
  estimation strategy profile $g$ to minimize $\bar J(g)$ given by
 \begin{equation} \label{eq:total-cost-infinite}
   \bar J(g) = \limsup_{T \rightarrow \infty} \frac{1}{T}J_{T}(g).
 \end{equation}
\end{problem}

As was the case for the estimation problem presented in
Sec.~\ref{sec:estimation}, a salient feature of the model is that
the estimates are generated using different information  while the objective
is to minimize a common coupled estimation error given
by~\eqref{eq:total-cost-finite} or~\eqref{eq:total-cost-infinite}. This
feature makes the Problems~\ref{prob:finite} and~\ref{prob:infinite}
conceptually different from the standard filtering problem of minimizing the
MMSE error.  

\begin{remark}
  For Problem~\ref{prob:finite}, the assumption that the
    dynamics, measurements, and cost are time-homogeneous is made simply for convenience of
    notation. As will be evident from the analysis, the results for
    Problem~\ref{prob:finite} generalize to the setting of time-varying
  dynamics, measurements, and cost in a natural manner.
\end{remark}

\subsection{Roadmap of the results} \label{sec:roadmap}

The main idea behind identifying a solution for Problem~\ref{prob:finite} is
as follows. We observe that the choice of the estimates only affects the
instantaneous estimation error but does not affect the evolution of the system
or the estimation error in the future. Therefore, the problem of choosing an
estimation profile $g = (g_1, \dots, g_n)$ to minimize $J_T(g)$ is equivalent
to solving the following $T$ separate optimization problems:
\begin{equation}
  	\min_{(g_{1,t}, \dots, g_{n,t})} \EXP[c(x(t), \hat z(t))], \quad
  	\forall t \in \{ 1, \dots, T \}.
    \label{eq:Pt}
    %\tag{$P_t$}
\end{equation}
Since the communication graph is strongly connected, the information $I_i(t)$
available at agent~$i$ can be written as $I^\com(t) \cup I^\loc_i(t)$, where 
\[
  I^\com(t) = \bigcap_{i \in N} I_i(t) = y(1:t-\DELAY^*)
\]
is the \emph{common information} among all agents (recall that $\DELAY^*$ is
the weighted diameter of the communication graph) and 
\[
  I^\loc_i(t) = I_i(t) \setminus I^\com(t)
\]
is the \emph{location information} at agent~$i$. Thus, we may view
Problem~\eqref{eq:Pt} as an estimation problem with $n$ agents where agents have
local and common information and, therefore, use the results of
Sec.~\ref{sec:estimation} to derive the MTMSE filtering strategy. 
To do so, we define variables which are equivalent to the auxiliary
variables defined in Sec.~\ref{sec:opt-est}:
\begin{itemize}
  \item All agents' \emph{common estimate} of state $x(t)$ given the common
    information $I^\com(t)$ at all agents. We denote this estimate by $\hat
    x^\com(t)$ and it is equal to $\EXP[ x(t) | I^\com(t) ]$.

  \item All agents' common estimate of the local information at agent~$i$
    given the common information. We denote this estimate by $\hat
    I^\loc_i(t)$ and it is equal to $\EXP[ I^\loc_i(t) | I^\com(t)]$. 
    
  \item The innovation in the local information at agent~$i$ with respect to
    the common information. We denote this innovation by $\tilde I_i(t)$ and
    it is equal to $I_i(t) - \hat I_i(t)$. 
\end{itemize}
Furthermore, we let $\hat \Theta_i(t)$ denote the covariance $\COV(x(t),
\tilde I_i(t))$ and $\hat \Sigma_{ij}(t)$ denote the covariance $\COV(\tilde
I^\loc_i(t), \tilde I^\loc_j(t))$. 

In order to use the results of Theorem~\ref{thm:estimation}, we need to derive
expressions for recursively updating the above variables and covariances, which
we do next. 

\subsection{Recursive expressions for auxiliary variables and covariances}

The information structure of the problem is effectively equal to
$\DELAY^*$-step delayed information structure~\cite{Witsenhausen1971}. 
To derive recursive expressions for auxiliary variables and covariances, we
follow the central idea of~\cite{Witsenhausen1971} and express the system
variables in terms of \emph{delayed state} $x(t - \DELAY^* + 1)$. 

\subsubsection{Delayed state estimates and common estimates}
We define
\begin{align}
  \hat x(t-\DELAY^*+1) &= \EXP[ x(t-\DELAY^*+1) \,|\, I^\com(t) ]
  \notag \\
  &= \EXP[ x(t-\DELAY^*+1) \,|\, y(1\COLON t-\DELAY^*) ]
\end{align}
as the \emph{delayed state estimate} of the state and let 
\[
  \tilde x(t - \DELAY^* +1) = x(t - \DELAY^* +1) - \hat x(t - \DELAY^* +1)
\]
denote the corresponding estimation error and $P(t - \DELAY^* + 1) =
\VAR(\tilde x(t-\DELAY^*+1))$ denote the estimation error covariance. Note
that $\hat x(t-\DELAY^*+1)$ is the one-step prediction estimate in centralized
Kalman filtering and can be updated as follows. Start with $\hat x(1)
= 0$ and for $t \ge 1$, update
\begin{align} 
  \hat x(t+1) &= A \hat x(t) + A K(t) [y(t) - C \hat x(t)],
  \label{eq:KF-estimate}
  \shortintertext{where}
  K(t) &= P(t)C^\TRANS [CP(t)C^\TRANS + R]^{-1}
  \label{eq:KF-gain}
\end{align}
is the Kalman gain. Furthermore, the error covariance $P(t)$ can be
pre-computed recursively using the forward Riccati equation: $P(1) = \Sigma_x$
and for $t \ge 1$,
\begin{equation} \label{eq:KF-riccati}
  P(t+1) = A \Delta(t) P(t)\Delta(t)^\TRANS A^\TRANS + AK(t)RK(t)^\TRANS A^\TRANS
  + Q,
\end{equation}
where $\Delta(t) = I - K(t) C$.

Now, observe that we can compute the common estimate $\hat x^\com(t)$ using a
$(\DELAY^*-1)$-step propagation of the delayed state estimate $\hat x(t - \DELAY^*
+ 1)$ as follows:
\begin{equation} \label{eq:xhat}
  \hat x^\com(t) = A^{\DELAY^* - 1} \hat x(t - \DELAY^* + 1).
\end{equation}

\subsubsection{Local estimates and local innovation}

To find a convenient expression for local innovation $\tilde I^\loc_i(t)$, we
express $I^\loc_i(t)$ in terms of the delayed state $x(t - \DELAY^* + 1)$.
For that matter, for any $t, \ell \in \mathbb{Z}_{>0}$, define the $d_x \times
1$ random vector $w^{(k)}(\ell,t)$ as follows:
\begin{equation}
  w^{(k)}(\ell, t) = \smashoperator[r]{\sum_{s=\max \{ 1,t-k \}}^{t-\ell-1}}
  A^{t-\ell-s-1} w(s),
  \label{eq:wbar-generic}
\end{equation}
where $w^{(k)} (\ell,t)$ is the weighted accumulated process noise from time $\max \{ 1,t-k \}$ to time $t-\ell-1$.
Note that $w^{(k)} (\ell,t) = 0$ if $t \le \min\{k, \ell + 1\}$ or $\ell \ge k$.
For any $t \ge k$, we may write
\begin{align} \label{eq:x-delayed}
  x(t) &= A^{k} x(t-k) + w^{(k)}(0,t), \\
  y_i(t) &= C_i A^k x(t-k) + C_i w^{(k)}(0,t) + v_i(t).
\end{align}

By definition $I^\loc_i(t) \subseteq y(t-\DELAY^*+1\COLON t)$. Thus, for any $i
\in N$, we can identify matrix $C^\loc_i$ and random vectors $w^\loc_i(t)$ and
$v^\loc_i(t)$ (which are linear functions of $w(t-\DELAY^*+1\COLON t-1)$ and
$v_i(t-\DELAY^*+1\COLON t)$) such that
\begin{equation}\label{eq:loc}
  I^\loc_i(t) = C^\loc_i x(t-\DELAY^*+1) + w^\loc_i(t) + v^\loc_i(t).
\end{equation}

As an example, we write the expressions for $(C_i^\loc, w^\loc_i(t), v^\loc_i(t))$ for the
delayed sharing and neighborhood sharing information structures below. 
For any $\ell \le \DELAY^*$, define
\begin{align*}
  \mathcal W_i(\ell, t) &= \VVEC( C_i w^{(\DELAY^*-1)}(\DELAY^*-1,t),
   % C_i w^{(\DELAY^*-1)}(\DELAY^*-2,t) ,
    \dots, 
  C_i w^{(\DELAY^*-1)}(\ell,t)),
  \\
  \mathcal C_i(\ell) &= \ROWS(C_i , C_i A , \dots , C_i A^{\DELAY^*-\ell-1}),
  \\
  \mathcal V_i(\ell, t) &= \VVEC( v_i(t-\DELAY^*+1), \dots, v_i(t-\ell ) ).
\end{align*}

\begin{example1c}
  For the $\DELAY$-step delayed sharing information structure
  $I^\loc_i(t) = y_i(t-\DELAY+1 \COLON t)$. Thus,
  \(
    C_i^\loc = \mathcal C_i(0)
  \),
  \(
    w_i^\loc(t) = \mathcal W_i(0, t)
  \), and
  \(
    v_i^\loc(t) = \mathcal V_i(0,t)
  \).
\end{example1c}
\begin{example2c}
  For the neighborhood sharing information structure,
	\(
	  I_i(t) = \bigcup_{k = 0}^{\DELAY^*} \bigcup_{j \in N^k_i} \{ y_j(1\COLON t-k) \}.
	\)
    Thus,
    \begin{align}
    C^\loc_i &=  \textstyle
    \ROWS\big( \bigodot_{\ell=0}^{\DELAY^*-1} \bigodot_{j \in N^\ell_i}
      \mathcal C_j(\ell)
    \big),
    \notag \\
    w^\loc_i(t) &=  \textstyle
    \VVEC\big( \bigodot_{\ell=0}^{\DELAY^*-1} \bigodot_{j \in N^\ell_i}
      \mathcal W_j(\ell,t)
    \big),
    \notag \\
    v^\loc_i(t) &= \textstyle \VVEC\big( \bigodot_{\ell=0}^{\DELAY^*-1} \bigodot_{j \in N^\ell_i}
      \mathcal V_j(\ell,t)
    \big).
    \tag*{$_\square$}
  \end{align}
\end{example2c}

Now, a key-result is the following.

\begin{lemma} \label{lemma:independence}
  $w^\loc_i(t)$, $v^\loc_i(t)$, $\tilde x(t-\DELAY^*+1)$, and $I^\com(t)$ are independent.
\end{lemma}

\begin{proof}
  Observe that $I^\com (t) = y(1 \COLON t-\DELAY^*)$ and ${\tilde
  x(t-\DELAY^*+1)}$ are functions of the primitive random variables up to time
  $t-\DELAY^*$, while $w^\loc_i(t)$ and $v^\loc_i(t)$ are functions of the
  primitive random variables from time ${t-\DELAY^*+1}$ onwards. Thus,
  $w^\loc_i(t)$ and $v^\loc_i(t)$ are independent of $\tilde x(t-\DELAY^*+1)$
  and $I^\com(t)$. Furthermore, (A3) implies that $w^\loc_i(t)$ and
  $v^\loc_i(t)$ are independent of each other. Note that $\tilde
  x(t-\DELAY^*+1)$ is the estimation error when estimating $x(t-\DELAY^*+1)$
  given $I^{\com}(t)$ and is, therefore, uncorrelated with $I^\com(t)$. Since all random variables are Gaussian, $\tilde x(t-\DELAY^*+1)$ and $I^\com(t)$
  being uncorrelated also means that they are independent.
\end{proof}

Combining Lemma~\ref{lemma:independence} with~\eqref{eq:loc}, we get
\begin{equation} \label{eq:hatIloc}
    \hat I^\loc_i(t) = \EXP[ I^\loc_i(t) | I^\com(t)] = 
    C^\loc_i \hat x(t - \DELAY^* + 1).
\end{equation}
Combining this with~\eqref{eq:loc}, we get,
\begin{equation} \label{eq:tildeIloc}
  \begin{split}
  \tilde I^\loc_i(t) &= I^\loc_i(t) - \hat I^\loc_i(t) \\ 
  &= C^\loc_i \tilde x(t - \DELAY^* + 1) + w^\loc_i(t) + v^\loc_i(t).
\end{split}
\end{equation} 

\subsubsection{Covariances}

Let $P^w_{ij}(t)$ denote $\COV(w^\loc_i(t), w^\loc_j(t))$ and $P^v_{ij}(t)$
denote $\COV(v^\loc_i(t), v^\loc_j(t))$. Note that these can be computed from
he expressions of $w^\loc_i(t)$ and $v^\loc_i(t)$, which were derived earlier
based on the communication graph. 

Eq.~\eqref{eq:tildeIloc} and Lemma~\ref{lemma:independence} imply that
\begin{equation} \label{eq:hatSigma}
  \begin{split}
    \hat \Sigma_{ij}(t) &= \COV(\tilde I^\loc_i(t), \tilde I^\loc_j(t)) \\
    &= 
    C^\loc_i P(t-\DELAY^*+1)C^\loc_j\strut^\TRANS + 
    P^w_{ij}(t) + P^v_{ij}(t),
  \end{split}
\end{equation}
where $P(t)$ is computed using~\eqref{eq:KF-riccati}.

Furthermore, Eqs.~\eqref{eq:x-delayed} and~\eqref{eq:tildeIloc} and
Lemma~\ref{lemma:independence} imply that
\begin{equation} \label{eq:hatTheta}
  \begin{split}
    \hat \Theta_i(t) &= \COV(x(t), \tilde I^\loc_i(t)) \\
    &= A^{\DELAY^* - 1}P(t - \DELAY^* + 1)C^\loc_i\strut^\TRANS + 
    P^\sigma_i(t),
  \end{split}
\end{equation}
where $P^\sigma_i(t) = \COV(w^{(\DELAY^*-1)}(0,t), w^\loc_i(t))$ and $P(t)$ is
computed using~\eqref{eq:KF-riccati}.

\subsection{Main result for Problem~\ref{prob:finite}}\label{sec:finite}

As mentioned in Sec.~\ref{sec:roadmap}, the problem of choosing the MTMSE
estimation strategy $g = (g_1, \dots, g_T)$ to minimize $J_T(g)$ is equivalent
to solving $T$ separate estimation sub-problems given by~\eqref{eq:Pt}. Based on
Theorem~\ref{thm:estimation}, the MTMSE estimate of each of these sub-problems
is given as follows.

\begin{theorem} \label{thm:finite}
  Under assumptions (A1)--(A3), the filtering strategy which minimizes the
  team mean-squared error in Problem~\ref{prob:finite} is a linear
  function of the measurements. Specifically, the MTMSE estimates at
  time~$t$ may be written as
  \begin{equation} \label{eq:filtering}
    \hat z_i(t) = L_i \hat x^\com(t) + F_i(t) \tilde I^\loc_i(t)
  \end{equation}
  where $\hat x^\com(t)$ and $\tilde I^\loc_i(t)$ are computed
  using~\eqref{eq:xhat} and~\eqref{eq:tildeIloc}. The gains $\{F_i(t)\}_{i \in
  N}$ satisfy the following system of matrix equations
  \begin{equation}\label{eq:matrix-eq}
    \sum_{j \in N} \Big[
      S_{ij} F_j(t) \hat \Sigma_{ji}(t) - S_{ij} L_j \hat \Theta_i(t)
    \Big] = 0,
    \quad \forall i \in N,
  \end{equation}
  where $\hat \Sigma_{ij}(t)$ and $\hat \Theta_i(t)$ are computed
  using~\eqref{eq:hatSigma} and~\eqref{eq:hatTheta}. Eq.~\eqref{eq:matrix-eq}
  has a unique solution which can be written as 
  \begin{equation} \label{eq:F}
    F(t) = \Gamma(t)^{-1} \eta(t),
  \end{equation}
  where
  \begin{align*} \label{eq:gains-main}
    F(t) &= \VVEC(F_1(t), \dots, F_n(t)), \\
    \eta(t) &= \VVEC(S_{1 \bullet} L\hat \Theta_1(t), \dots, S_{n \bullet} L \hat \Theta_n(t)), \\
    %S_i &= [S_{i1}, \dots, S_{in}], \\
    \Gamma(t) &= [\Gamma_{ij}(t)]_{i,j \in N},
    \quad\text{where }
    \Gamma_{ij}(t) = \hat \Sigma_{ij}(t) \otimes S_{ij}.
  \end{align*}

  Furthermore, the minimum team mean-squared error is given by 
  \begin{equation}
    J^*_T = \sum_{t = 1}^{T} \big[\Tr(L^\TRANS S L P_0(t) ) -\eta(t)^\TRANS
    \Gamma(t)^{-1} \eta(t)  \big],
  \end{equation}
  where $P_0(t) = \VAR({x(t)- \hat x^\com(t)})$ and is given by
    \begin{align} %\label{eq:hat-P}
      P_0(t) &= A^{\DELAY^*-1} P(t-\DELAY+1) (A^{\DELAY^*-1})^\TRANS
      + \Sigma^w(t),
      \label{eq:P-0}
    \end{align}
    and $\Sigma^w(t) = \VAR(w^{(\DELAY^*-1)}(0,t))$.
\end{theorem}

\begin{proof}
  The expressions for the MTMSE estimates~\eqref{eq:filtering} and the
  corresponding gains~\eqref{eq:matrix-eq} follow immediately from
  Theorem~\ref{thm:estimation}. Now, since $R_{ii}$ is positive
  definite (which is part of (A2)), standard results from Kalman
  filtering~\cite[Section 3.4]{Caines1987} imply that $P(t)$ is positive
  definite. Using this fact in~\eqref{eq:hatSigma} implies that $\hat
  \Sigma_{ii}(t)$ is positive definite. Therefore, the vectorized
  formula~\eqref{eq:F} follows from Lemma~\ref{lemma:sandell}.

  The expression for the minimum team mean-squared error follow from
  an argument similar to that in the proof of Theorem~\ref{thm:estimation}.
  The expression for $P_0(t)$ follows from~\eqref{eq:xhat}
  and~\eqref{eq:x-delayed}.
\end{proof}

\begin{remark}
   Remark~\ref{rem:structure} about the structure of the MTMSE estimates
   continues to hold for filtering setup as well. The first term in the MTMSE
   estimate~\eqref{eq:filtering} is the MMSE estimate of the current state
   based on the common information. The second term is a ``correction'' which
   depends on the innovation in the local measurements. 
\end{remark}

\begin{remark}
   As in the estimation setup, the gains which multiply the innovation
   in~\eqref{eq:filtering} are coupled and depend on the weight matrix~$S$. 
\end{remark}

\begin{remark} \label{rem:time}
   Since we have assumed that the dynamics are time-homogeneous, the processes
   $\{w^{(\DELAY^*-1)}(0,t)\}_{t \geq \DELAY^*}$, $\{w^\loc_i(t)\}_{t \geq
   \DELAY^*}$, and $\{v^\loc_i(t)\}_{t \geq \DELAY*}$ are stationary. Hence,
   for $t \geq \DELAY^*$, the covariance matrices $\Sigma^w(t)$,
   $P^\sigma_i(t)$, $P^w_{ij}(t)$, and $P^v_{ij}(t)$ are constant.
\end{remark}

\begin{remark}\label{rem:sparse}
  Note that $\hat
  \Sigma_{ij} \otimes S_{ij} = \mathbf{0}$ when $S_{ij} = 0$. Therefore, 
  when the weight matrix $S$ is sparse, as is the case for the
  cost~\eqref{eq:platoon-cost}, $\hat \Sigma_{ij}$ (and, therefore,
  $P^w_{ij}(t)$ and $P^v_{ij}(t)$) need to computed only for
  those $i, j \in N$ for which $S_{ij} \neq \mathbf{0}$. 
\end{remark}

% \begin{remark}
% \textcolor{blue}{Matrices being time dependent does not change the results for Theorem 1 as it is a static problem.
% For Problem 2, the results regarding the common part of the information remains the same as
% this is a standard Kalman filtering equation and holds for the case when the matrices are time
% dependent. The results remain the same for equations (43) and (44) when the system matrices
% are time dependent as well.}
% \end{remark}

\subsection{Main result for Problem~\ref{prob:infinite}}\label{sec:infinite}
Now, we consider the infinite horizon MTMSE filtering
introduced in Problem~\ref{prob:infinite}, which can be thought of as a
``steady-state'' version of Sec.~\ref{sec:finite}. We first state a standard result from centralized Kalman
filtering~\cite{Caines1987}.
\begin{lemma} \label{lemma:riccati-convergence}
  Under (A2)--(A5), for any initial covariance $\Sigma_x \geq 0$, 
  the sequence $\{P(t)\}_{t \ge 1}$ given by~\eqref{eq:KF-gain} is 
  weakly increasing and bounded (in the sense of positive semi-definiteness). 
  Thus it has a limit, which we denote by $\bar P$. Furthermore,
  \begin{enumerate}
    \item $\bar P$ does not depend on $\Sigma_x$.
    \item $\bar P$ is positive semi-definite.
    \item $\bar P$ is the unique solution to the following algebraic Riccati equation.
    \begin{equation} \label{eq:P-t-ARE}
     \bar P = A \Delta \bar P \Delta^\TRANS A^\TRANS \\
    + A \bar K R \bar K^\TRANS A^\TRANS + Q,
  \end{equation}
  where $\bar K = \bar P C^\TRANS\big[ C \bar P C^\TRANS + R \big]^{-1}$ and $\Delta = I - \bar K C$.
  \vskip 4pt
  \item The matrix $(A - \bar K C)$ is asymptotically stable.
  \end{enumerate}
\end{lemma}

Recall from Remark~\ref{rem:time} that
$\Sigma^w(t)$, $P^\sigma_i(t), P^w_{ij}(t)$ and $P^v_{ij}(t)$ are constants
for $t \geq \DELAY^*$. We denote the corresponding values for $t \geq \DELAY^*$ as $\bar
\Sigma^w$, $\bar P^\sigma_i, \bar P^w_{ij}$, and $\bar P^v_{ij}$.
Now define:
\begin{align}
  \bar P_0 &= A^{\DELAY^*-1} \bar P (A^{\DELAY^*-1})^\TRANS + \bar \Sigma^w, 
  \label{eq:barP0}\\
  \bar \Sigma_{ij} &= C_i^{\loc} \bar P {C_j^{\loc}}^\TRANS + \bar P_{ij}^w +
\bar P_{ij}^v, 
  \label{eq:barSigma}  \\
  \bar \Theta_i &= A^{\DELAY^* -1 } \bar P C^\loc_i\strut^\TRANS + \bar P^\sigma_i.
  \label{eq:barTheta}
\end{align}

\begin{lemma} \label{lemma:covariance-inf}
Under (A2)--(A5), we have the following:
\begin{enumerate}
  \item $\lim_{t \rightarrow \infty} P_0(t) = \bar P_0$.
  \item $\lim_{t \rightarrow \infty} \hat  \Sigma_{ij}(t) = \bar \Sigma_{ij}$.
  \item $\lim_{t \rightarrow \infty} \hat \Theta_i(t) = \bar \Theta_i.$
\end{enumerate}
\end{lemma}

\begin{proof}
  All relations follow immediately from Lemma~\ref{lemma:riccati-convergence} and 
  Remark~\ref{rem:time}.
\end{proof}

\begin{theorem} \label{thm:strategy-inf}
  Under (A1)--(A5), the following time-homogeneous filtering strategy minimizes the
   team mean-squared error for Problem~\ref{prob:infinite}:
  \begin{equation} \label{eq:strategy-main-inf}
    \hat z_i(t) = L_i \hat x^\com(t) + \bar F_i \tilde I^\loc_i(t) ,
  \end{equation}
  where $\hat x^\com(t) = A^{\DELAY^* - 1} \hat x(t - \DELAY^* + 1)$ (which is same as~\eqref{eq:xhat}), $\hat x(t)$ is updated using the steady state version of~\eqref{eq:KF-estimate} given by  
\begin{equation} 
  \hat x(t+1) = A \hat x(t) + A \bar K [y(t) - C \hat x(t)],
  \label{eq:KF-estimate-inf}
\end{equation}  
and the gains $\{ \bar F_i \}_{i \in N}$ satisfy the following system of
  matrix equations:
  \begin{equation}\label{eq:matrix-eq-inf}
    \sum_{j \in N} \Big[
      S_{ij} \bar F_j \bar{\Sigma}_{ji} - S_{ij} L_j \bar{\Theta}_i
    \Big] = 0,
    \quad \forall i \in N,
  \end{equation}
  where $\bar \Sigma_{ij}$ and $\bar \Theta_i$ are given
  by~\eqref{eq:barSigma} and~\eqref{eq:barTheta}. 
  Eq.~\eqref{eq:matrix-eq-inf} has a unique solution and can be written more
  compactly as
  \begin{equation}\label{eq:F-inf}
    \bar F = \bar \Gamma^{-1} \bar \eta,
  \end{equation}
  where
  \begin{align*} \label{eq:gains-main-inf}
    \bar F &= \VVEC(\bar F_1, \dots, \bar F_n), \\
    \bar \eta &= \VVEC(S_{1 \bullet} L\bar{\Theta}_1, \dots, S_{n \bullet} L \bar{\Theta}_n), \\
    %S_i &= [S_{i1}, \dots, S_{in}], \\
    \bar \Gamma(t) &= [\bar \Gamma_{ij}]_{i,j \in N},
    \quad\text{where }
    \bar \Gamma_{ij} = \bar{\Sigma}_{ij} \otimes S_{ij}.
  \end{align*}
  Furthermore, the optimal performance is given by
  \begin{equation} \label{eq:optimal-total-cost-inf}
    J^* =
    \Tr(L^\TRANS S L \bar P_0 )
    -\bar \eta^\TRANS \bar \Gamma^{-1} \bar \eta,
  \end{equation}
  where $\bar P_0$ is given by~\eqref{eq:barP0}.
\end{theorem}

The proof of Theorem~\ref{thm:strategy-inf} is presented in Appendix~\ref{sec:inf-proof}.
%==================================================================

\section{Some illustrative examples}\label{sec:examples}

In this section, we present a few examples to illustrate the details of the
main results.

\subsection{Team mean-squared estimation in a UAV formation}
\label{sec:UAV}

\begin{figure}[tb]
  \centering
    \includegraphics{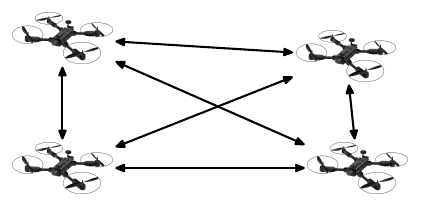}
    \caption{A four agent UAV formation. The arrows indicate communication
    links between the agents. Each link has delay~$2$.}
    \label{fig:UAV}
\end{figure}

Consider a UAV formation with $n$ agents as shown in 
Fig.~\ref{fig:UAV}.
Let $N = \{1,\dots,n\}$ and $x_i(t)$ denote the
state of agent~$i \in N$. For the ease of exposition, we assume that $x_i(t) \in
\reals$, which could correspond to say the altitude of the UAV. Let $x(t) =
\VVEC(x_1(t), \dots, x_n(t))$ denote the state of the system, which evolves as
\[
  x(t+1) = Ax(t) + w(t),
\]
where $A$ is a known $n \times n$ matrix and $w(t) \sim \N(0, Q)$. The agent~$i$ observes the state with noise, i.e., 
\[
  y_i(t) = C_{i} x(t) + v_i(t), \quad i \in N,
\]
where $v_i(t) \sim \N(0,R_i)$. 

The communication graph is as shown in Fig.~\ref{fig:UAV}, where each link is
assumed to have delay~2. Thus, the
information structure is given by
\[
  I_i(t) = \{ y(1 \COLON t-2), y_i(t-1{:}t)\}.
\]
The objective is to
determine the MTMSE filtering for per-step estimation error given
by~\eqref{eq:first-cost}, i.e., the agents want to estimate their local state
and ensure that the average of the local state estimates is close to the
average of their actual states. 

We first show the computations of the MTMSE estimates. Observe
that $I^\com(t) = y(1\COLON t-2)$ and 
\[
I^\loc_i(t) = \{ y_{i}(t-1),  y_i(t)\}.
\]
Thus,
\(
C^\loc_i =  \ROWS( C_{i} , C_i A),
\)
and
\begin{align*}
  w^\loc_i(t) &=  \VVEC( 0 , C_i w(t-1))
  , \quad 
  v^\loc_i(t) = \VVEC(v_i (t-1) , v_i(t)).
\end{align*}
As argued in Remark~\ref{rem:time}, the covariance matrices $\Sigma^w(t)$,
$P^\sigma_i(t)$, $P^w_{ij}(t)$, and $P^v_{ij}(t)$ are constant for $t \ge
\DELAY^*$. Thus, we only need to compute these for $t=1$ and $t \ge 2$.
Note that the weight matrix $S$ is dense, so we do not get the computational
savings described in Remark~\ref{rem:sparse}.

We have the following:
\begin{itemize}
  \item $\Sigma^w(1) = 0$ and for $t \ge 2$, $\Sigma^w(t) = Q$. 
  \item $P^\sigma_i(1) = \MATRIX{\mathbf{0}_{4 \times 1} & \mathbf{0}_{4 \times 1} }$ and for $t \ge 2$, 
    $P^\sigma_i(t) = \MATRIX{\mathbf{0}_{4 \times 1} & Q C^\TRANS}$.
  \item $P^w_{ij}(1) = \diag(0,0)$ and for $t \ge 2$,
    $P^w_{ij}(t) = \diag(0, C_i Q C_j^\TRANS)$. 
  \item $P^v_{ii}(1) = \diag(0,R_i)$ and $P^v_{ii}(t) = \diag(R_{i},
    R_i)$.
  \item $P^v_{ij}(t) = \diag(0,0)$ for $j \neq i$ and all $t$.
\end{itemize}
Substituting these, we get that $\hat \Sigma_{ij}(1) = \delta_{ij}\diag(0,R_i)$ and for
$t \ge 2$,
\begin{align*}
  \hat \Sigma_{ij}(t) = \MATRIX{C_i \\ C_i A} P(t-1) 
  \MATRIX{C_j \\ C_j A}^\TRANS + \MATRIX{\delta_{ij} R_i & 0
  \\ 0 & Q_{ij} + \delta_{ij} R_i}.
\end{align*}
Substituting these in~\eqref{eq:matrix-eq} or~\eqref{eq:F} gives us the
optimal gains. The MTMSE estimates can then be computed
using~\eqref{eq:filtering} as described in Sec.~\ref{sec:implementation}.

\begin{figure}[!t]
  \centering
  \begin{subfigure}[t]{1\linewidth} \centering
    \includegraphics[scale=0.95]{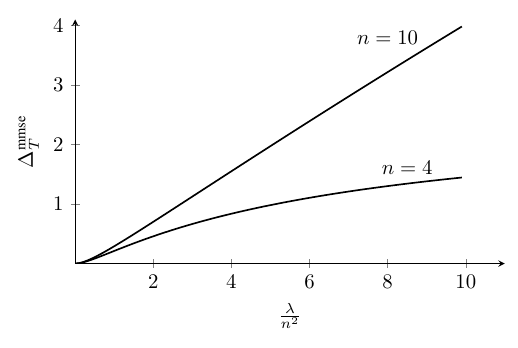}
    \caption{}
  \end{subfigure} \hfill
  \begin{subfigure}[t]{1\linewidth} \centering
    \includegraphics[scale=0.95]{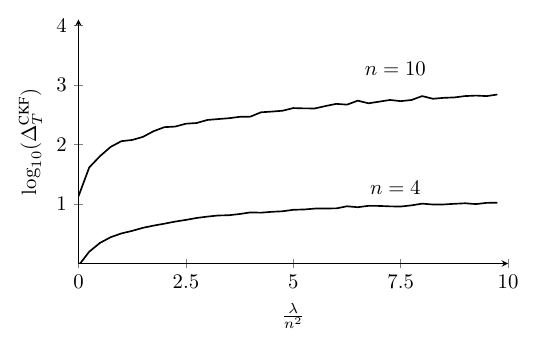}
    \caption{}
  \end{subfigure} \hfill
  \caption{Relative improvement of MTMSE filtering compared to (a)~MMSE strategy 
  for $4$ and $10$ number of agents, and (b)~consensus Kalman filtering (shown on a log scale) for UAV formation.}
  \label{fig:UAV-plots}
\end{figure}

We compare the performance of MTMSE filtering strategy with two baselines.
The first is MMSE strategy where, each agent ignores the cost coupling and
simply generates the MMSE estimates using
\begin{equation}
  \hat z^\mmse_i(t) = L_i \EXP[ x(t)|I_i(t)].
\end{equation}
It can be shown that performance of the MMSE
strategy is
\begin{multline}  \label{eq:cost-nonopt}
  J^\mmse_T = \Tr( L^\TRANS S L P_0(t)) \\
  + \sum_{i \in N} \Tr\bigg( K_i(t)^\TRANS L_i^\TRANS \sum_{j \in N} S_{ij} L_j \Big[
   K_j(t) \hat \Sigma_{ji}(t) - 2 \hat \Theta_{i}(t) \Big] \bigg).
\end{multline}
Recall that for this particular example we have $L = \mathbf{I}$. 

The second is a consensus based Kalman filter as described in~\cite{Olfati-Saber2007}.
We do not have a closed form expression for the weighted mean square
error of the consensus Kalman filter, so we evaluate the performance 
$J^{\text{CKF}}_T$ using Monte Carlo evaluation 
averaged over $1000$ sample paths.

For the numerical experiments we pick
\[
A_{ij} = \begin{cases} 0.65, & \quad i = j \\ 0.1, & \quad \text{elsewhere} \end{cases}
\]
$C_1 = 2 \times \textbf{1}_{1 \times n}$, and  for $i \neq 1$, 
$C_i = 0.1 e_i$, where $e_i$ is a vector with only the $i_{th}$ element equal to one and the rest zero, $Q=\mathbf{I}, R=0.1\mathbf{I}$, and $T=100$.

The relative improvements 
\[
  \Delta^\mmse_T = \frac{J^\mmse_T - J^*_T}{J^*_T}
   \quad \text{and} \quad
  \Delta^{\text{CKF}}_T = \frac{J^{\text{CKF}}_T - J^*_T}{J^*_T} 
\]
of the MTMSE strategy compared to MMSE strategy and consensus Kalman
filtering as a function of $\lambda$ are shown in Fig.~\ref{fig:UAV-plots}.
These plots show that the MTMSE strategy outperforms the MMSE and
consensus Kalman filtering strategies by up to a factor of 4 and 600
in the relative improvements for $n=10$ and $\frac{\lambda}{n^2}=10$. This improvement
in performance will increase with the number of agents. 

\subsection{Team mean-squared estimation in a vehicular platoon}
\label{sec:platoon}

\begin{figure}[!t]
  \centering
    \includegraphics{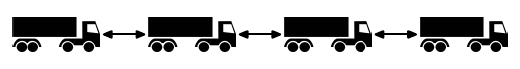}
    \caption{A four agent vehicular platoon. The arrows indicate communication
    links between the agents.}
    \label{fig:platoon}
\end{figure}

Now we consider a vehicular platoon with four agents shown in
Fig.~\ref{fig:platoon}. As before, let $x_i(t) \in \reals$ denote the position
of the platoon. We assume that the dynamics and the observation model are
similar to that described in Sec.~\ref{sec:UAV} (but with different $A$ and $C$
matrices). 

The communication graph is as shown in Fig.~\ref{fig:platoon}. Thus, the
information structure is given by
\begin{align*}
  I_1(t) &= \{ y_1(1 \COLON t), y_2(1 \COLON t-1), y_3(1 \COLON t-2), y_4(1 \COLON t-3) \},
  \\
  I_2(t) &= \{ y_1(1 \COLON t-1), y_2(1 \COLON t), y_3(1 \COLON t-1), y_4(1 \COLON t-2) \},
  \\
  I_3(t) &= \{ y_1(1 \COLON t-2), y_2(1 \COLON t-1), y_3(1 \COLON t), y_4(1 \COLON t-1) \},
  \\
  I_4(t) &= \{ y_1(1 \COLON t-3), y_2(1 \COLON t-2), y_3(1 \COLON t-1), y_4(1 \COLON t) \}.
\end{align*}
The objective is to determine the MTMSE filtering for per-step estimation
error given by~\eqref{eq:platoon-cost}, i.e., the agents want to estimate
their local states and ensure that the difference between the estimates of
adjacent agents is close to difference between their actual states.

We first show the computations of the MTMSE estimates. Observe
that $I^\com(t) = y(1\COLON t-3)$ and 
\begin{align*}
  I^\loc_1(t) &= \{ y_1(t-2 \COLON t), y_2(t-2 \COLON t-1), y_3(t-2) \},
  \\
  I^\loc_2(t) &= \{ y_1(t-2 \COLON t-1), y_2(t-2 \COLON t), y_3(t-2 \COLON t-1), \\
  & \hspace{6 cm} y_4(t-2) \},
  \\
  I^\loc_3(t) &= \{ y_1(t-2), y_2(t-2 \COLON t-1), y_3(t-2 \COLON t), \\
  & \hspace{5.25 cm} y_4(t-2 \COLON t-1) \},
  \\
  I^\loc_4(t) &= \{ y_2(t-2), y_3(t-2 \COLON t-1), y_4(t-2 \COLON t) \}.
\end{align*}

Similar to the previous example, the covariance matrices $\Sigma^w(t)$,
$P^\sigma_i(t)$, $P^w_{ij}(t)$, and $P^v_{ij}(t)$ are constant for $t \ge
\tau^*$. Thus, we need to compute these for $t = 1$, $t=2$, and $t \ge 3$. In
addition, since the cost matrix $S$ is sparse, we only need to compute
$P^w_{ij}(t)$ and $P^v_{ij}(t)$ for $j \in \{i-1, i, i+1\} \cap N$ (see
Remark~\ref{rem:sparse}). The details for computing $\hat \Sigma_{ij}$ are
similar to the previous section and are omitted due to space limitations. The
MTMSE estimates can be computed using~\eqref{eq:filtering} as described in
Sec.~\ref{sec:implementation}.

\begin{figure}[!t]
  \begin{subfigure}[t]{1\linewidth} \centering
    \includegraphics[scale=0.95]{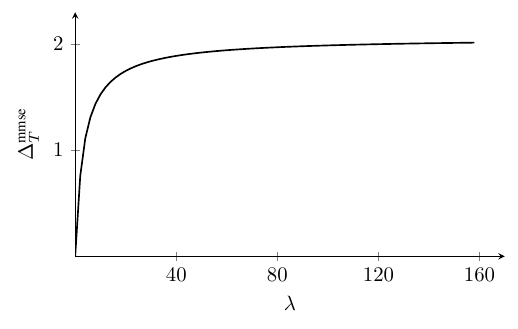}
    \caption{}
  \end{subfigure} \hfill
  \begin{subfigure}[t]{1\linewidth} \centering
    \includegraphics[scale=0.95]{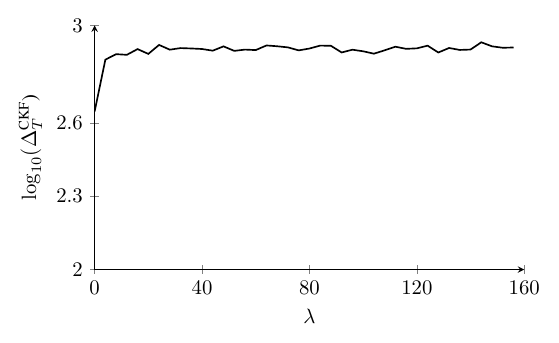}
    \caption{}
  \end{subfigure} \hfill
  \caption{Relative improvement of MTMSE filtering compared to (a)~MMSE strategy and (b)~consensus Kalman filtering (shown on a log scale) for vehicular platoon.}
  \label{fig:platoon-plots}
\end{figure}

We compare the performance of MTMSE filtering strategy with the MMSE strategy and the consensus Kalman filtering as before.

For the numerical experiment in this part, we pick
\[
A = \begin{bmatrix}
0.9 & 0 & 0 & 0\\ 0.7 & 0.9 & 0 & 0\\ 0.7 & 0.7 & 0.9 & 0
\\0.5 & 0.7 & 0.7 & 0.9 
\end{bmatrix},
\]
$C_i = \textbf{I}_n$, $Q=\mathbf{I}, R=0.1\mathbf{I}$, and $T=100$.

The relative improvements as a
function of $\lambda$ are shown in Fig.~\ref{fig:platoon-plots}. 
These plots show that the MTMSE strategy outperforms the MMSE and
consensus Kalman filtering strategies by up to a factor of 2 
and 800. Again, this improvement
in performance will increase with the number of agents. 

\section{Discussion of the results} \label{sec:discussion}

\subsection{Implementation of MTMSE filtering strategy} \label{sec:implementation}
In this section, we provide the details about implementing the MTMSE filtering strategies for both the finite and infinite horizon setups. 

\subsubsection{Implementation of finite horizon MTMSE filtering strategy}

Based on Theorem~\ref{thm:finite}, the MTMSE filtering strategy can be
implemented as follows.

\paragraph{Computing the gains}
The gains $\{F(t)\}_{t=1}^T$ are computed offline as follows.
First the variance $\{P(t)\}_{t=1}^T$ are
computed using the forward Riccati equation~\eqref{eq:KF-riccati}.
Then, the covariances $\{\hat \Sigma_{ij}(t)\}_{t=1}^T$ and $\{\hat
\Theta_i(t)\}_{t=1}^T$ are computed for all $i,j \in N$. 
Thereafter, the gains $\{K(t)\}_{t=1}^T$ are computed using~\eqref{eq:KF-gain}
and the gains $\{F(t)\}_{t=1}^T$ are computed using~\eqref{eq:F}.

Finally, the gains $\{K(t)\}_{t=1}^T$ and $\{F_i(t)\}_{t=1}^T$ are stored in
agent~$i$.

\paragraph{Computing the MTMSE estimates}

Agent $i \in N$ carries out the following computations to generate $\hat
z_i(t)$. First, it computes the delayed centralized estimate $\hat x(t -
\DELAY^* + 1)$ using~\eqref{eq:KF-estimate}. Then, it uses $\hat x(t -
\DELAY^* + 1)$ to compute $\hat x^\com(t)$ and $\hat I^\loc_i(t)$
using~\eqref{eq:xhat} and~\eqref{eq:hatIloc}, respectively. Then, it uses
$\hat x^\com(t)$ and $I^\loc_i(t)$ to generate the MTMSE estimate as follows
\[
  \hat z_i(t) = L_i \hat x^\com(t) + F_i(t)( I^\loc_i(t) - \hat I^\loc_i(t) ).
\]

\subsubsection{Implementation of infinite horizon MTMSE filtering strategy}

Based on Theorem~\ref{thm:strategy-inf}, the MTMSE filtering strategy can be
implemented as follows.

\paragraph{Computing the gains}

The gains $\{\bar F_i \}$ are computed offline as follows.
First the variance $\bar P$ is
computed using the forward algebraic Riccati equation~\eqref{eq:P-t-ARE}.
Then, the covariances $\bar P_0$, $\bar \Sigma_{ij}$, and $\bar \Theta_i$ 
are computed for all $i,j \in N$ using~\eqref{eq:barP0}-\eqref{eq:barTheta}. 
Thereafter, the gain $\bar K$ is computed using Lemma~\ref{lemma:riccati-convergence} and the gain $\bar F$ is computed using~\eqref{eq:F-inf}. Finally, the gains $\bar K$ and $\bar F$ are stored in
agent~$i$.

\paragraph{Computing the MTMSE estimates}

Agent $i \in N$ carries out the following computations to generate $\hat
z_i(t)$. First, it computes the delayed centralized estimate $\hat x(t -
\DELAY^* + 1)$ using~\eqref{eq:KF-estimate-inf}. Then, it uses $\hat x(t -
\DELAY^* + 1)$ to compute $\hat x^\com(t)$ and $\hat I^\loc_i(t)$
using~\eqref{eq:xhat} and~\eqref{eq:hatIloc}, respectively. Then, it uses
$\hat x^\com(t)$ and $I^\loc_i(t)$ to generate the MTMSE estimate as follows
\[
  \hat z_i(t) = L_i \hat x^\com(t) + \bar F_i( I^\loc_i(t) - \hat I^\loc_i(t) ).
\]

\subsection{Connection to decentralized stochastic control} \label{sec:disc-con}

One of the most celebrated results in centralized stochastic control of 
linear systems with quadratic cost and Gaussian disturbance (so-called 
LQG setup) is the separation of estimation and control. In particular, the
optimal control action is equal to a gain multiplied by the current state 
estimate. The computation of the gain matrix and the estimate are separated 
from each other. The gain matrix is computed based on the solution of a 
backward Riccati equation where the state estimates are updated based on the 
Kalman filtering equation (which is a forward Riccati equation). The forward
and the backward Riccati equations are decoupled and can be solved separately.

These simplifications do not hold for decentralized control of LQG systems. 
In general, non-linear strategies may outperform the best linear strategies.
Linear strategies are known to be optimal only for specific models~\cite{Ho1972,Yoshikawa1975,Asghari2018,Afshari2020,Swigart2010,Nayyar2018}. But in these cases there is no separation of estimation 
and control. 

The results of this paper shed light on the lack of separation in decentralized
control of LQG systems. We explain this in Appendix~\ref{app:one-step-delayed} using the example of decentralized stochastic control with 
one-step delayed information structure~\cite{Kurtaran1974,Sandell1974,Yoshikawa1975}. For this model, 
we show that the decentralized control problem is equivalent to a MTMSE 
filtering problem, where the weight matrix depends on the solution of a backward Riccati equation. 
As shown in Theorem~\ref{thm:finite}, the gains for MTMSE filtering 
depends on the weight matrix $S$ in the cost function. That is the reason 
that the computation of the state estimate is not separated from the 
computation of the controller gains. 

\subsection{Trade-off between filter complexity and estimation accuracy} \label{sec:disc-trade}

For graphs with neighborhood sharing information 
structure, the dimension of $\tilde I_i^\loc(t)$ and $F_i(t)$ are 
proportional to the diameter $\tau^*$ of the graph. It is possible to 
trade-off the implementation complexity with the filtering accuracy by
``shedding'' information at each agent. We explain this via the 
example of Sec.~\ref{sec:platoon}. 

We consider two approximate information structures for this example, which we
denote by $\{I^{(1)}_i(t) \}_{i \in N}$ and $\{I^{(2)}_i(t) \}_{i \in N}$. For
both these information structures, the common information is the same as
before, i.e., 
\[
  I^{\com, (m)}(t) \DEFINED \bigcap_{i \in N} I^{(m)}_i(t) = y(1\COLON{t-3}), 
  \quad m \in \{1, 2\}.
\]
But the local information $I^{\loc,(m)}_i(t) \DEFINED I^{(m)}_i(t) \setminus
I^{\com,(m)}(t)$ is a subset of the original $I^\loc_i(t)$. In particular, we
assume the following.
\begin{enumerate}
  \item \textbf{IS\textsubscript{1}}: In the first approximation, each agent just uses the measurements
    from a time window of size two to ``correct'' the common information based
    estimate, i.e.,
    \begin{align*}
      I^{\loc,(1)}_1(t) &= \{ y_1(t-1 \COLON t), y_2(t-1)\},
      \\
      I^{\loc,(1)}_2(t) &= \{ y_1(t-1), y_2(t-1 \COLON t), y_3(t-1) \},
      \\
      I^{\loc,(1)}_3(t) &= \{ y_2(t-1), y_3(t-1 \COLON t), y_4(t-1) \},
      \\
      I^{\loc,(1)}_4(t) &= \{ y_3(t-1), y_4(t-1 \COLON t) \},
    \end{align*}
  \item \textbf{IS\textsubscript{2}}: In the second approximation, each agent
    justs uses its local measurements to ``correct'' the common information
    based estimate, i.e., 
    \[
      I^{\loc,(2)}_i(t) = y_i({t-2}\COLON t).
    \]
\end{enumerate}
For completeness, we refer to the original information structure as
IS\textsubscript{0}. 
Note that $I^{\loc,(m)}_i(t) \subset I^\loc_i(t)$, therefore any filtering
strategy based on the approximate information structure $\{I^{(m)}_i(t)\}_{i
\in N}$ can be implemented in the original information structure $\{I_i(t)
\}_{i \in N}$. The size of $I^{\loc}_i(t)$ (and therefore $\tilde
I^\loc_i(t)$) for the different information structures is shown in
Table~\ref{tab:compare}.

To compare the peformance of these three information structures, we note that
the structure of the weight matrix $S$ implies that $\lim_{\lambda \to \infty}
J^*_T/\lambda$ is a constant. So, we evaluate $J^*_T/\lambda$ for large value
of $\lambda$ ($\lambda = 100$) and compare the performance of the three
information structures. The results are also shown in Table~\ref{tab:compare}.
\begin{table}[!th]
  \caption{Comparison of the size and performance of the three information
  structures for the values of parameters of Sec.~\ref{sec:platoon} and
$\lambda = 100$.}
  \label{tab:compare}
  \begin{tabular}{r@{\hskip 0.25em:\hskip0.25em}lccc}
    \toprule
    \multicolumn{2}{c}{\thead{Info structure}} & \multicolumn{2}{c}{\thead{Dimension of local
    info}} &
    \thead{Performance $J^*_T/\lambda$} 
    \\
    \cmidrule{3-4}
    \multicolumn{2}{c}{} & $i \in \{1, 4\}$ & $i \in \{2,3 \}$ & 
    \\
    \midrule
    IS\textsubscript{0} & $\{ I_i(t) \}_{i \in N}$ & 6 & 8 & 180.46  \\[2pt]
    IS\textsubscript{1} & $\{ I^{(1)}_i(t) \}_{i \in N}$ & 3 & 4 & 193.72  \\
    IS\textsubscript{2} & $\{ I^{(2)}_i(t) \}_{i \in N}$ & 3 & 3 & 252.09  \\[2pt]
    \bottomrule
  \end{tabular}
\end{table}

This example shows that it is possible to trade-off the complexity of the 
MTMSE filter with the estimation accuracy. Note that although the two approximate information structures are almost of
the same size, IS\textsubscript{1} has better performance than
IS\textsubscript{2}.
This is because IS\textsubscript{1} uses some
local infomration from the neighborhood nodes, while IS\textsubscript{2}
does not. This suggested that it is better to have some information from 
many agents rather than a lot of information from a few agents but a more 
detailed investigation is needed to quantify such a comparison.

\section{Conclusion}

In this paper, we investigate multi-agent estimation and filtering to minimize
team mean-square error. We show that the MTMSE estimates are given by
\[
  \hat z_i(t) = L_i \hat x^\com(t) + F_i(t) (I^\loc_i(t) - \hat I^\loc_i(t)).
\]
The first term of the estimate is the conditional mean of the current state
given the common information. The second term may be viewed as a
``correction'' which depends on the ``innovation'' in the local measurements.
A salient feature of this result is that the gains $\{F_i(t)\}_{i \in N}$
depend on the weight matrix~$S$. Using illustrative examples, we show that the
MTMSE estimates significantly smaller team mean-squared error as compared to
MMSE strategy and consensus Kalman filtering.

The results were derived under the assumptions that the state process
$\{x(t)\}_{t \ge 1}$ is a linear stochastic process and the observation
channels are linear and additive Gaussian noise. In future, we plan to
investigate team estimation of general stochastic processes over general
measurement channels, which will give rise to non-linear filtering equations.

Finally, our focus in this paper was to establish the structure of MTMSE
filtering and filtering strategies. Having identified this structure, it is
possible to implement the policy efficiently in a distributed manner. For
example, for the infinite horizon setup, it is possible to use a consensus Kalman
filter~\cite{Olfati-Saber2007,
Olfati-Saber2009,Kar2009,Cattivelli2010,Olfati-Saber2012,Battistelli2015}
 to keep track of the delayed state estimate $\hat x(t - \DELAY^* + 1)$
and use distributed algorithms to solve the linear system of equations $\bar
\Gamma \bar F = \bar \eta$ using distributed algorithms~\cite{Liu2017,
Yang2020, Wang2019}.
%+++++++++++++++++++++++++++++++

\appendices

\section{Proof of Theorem~\ref{thm:estimation}} \label{sec:one-step-proof}

\subsection{A preliminary result}
\begin{comment}
We start by stating basic properties of Gaussian random variables.
\begin{lemma}\label{lem:basic}
  Let $a$ and $b$ be jointly Gaussian zero-mean random variables with
  covariance $\left[\begin{smallmatrix} \Sigma_{aa} & \Sigma_{ab} \\
\Sigma_{ba} & \Sigma_{bb} \end{smallmatrix} \right]$. Then,
  \begin{enumerate}
    \item $\EXP[ a | b] = \Sigma_{ab} \Sigma_{bb}^{-1} b$
    \item For matrices $A$ and $B$ of appropriate dimensions,
      \[
        \EXP[ a^T A^T B b] = \Tr(A \Sigma_{ab} B^\TRANS)
        = \Tr(\Sigma_{ba} A^\TRANS B).
      \]
  \end{enumerate}
\end{lemma}
\end{comment}

In order to compute the gains and the performance, we need to compute
 $\hat \Theta_i = \COV(x,\tilde y_i)$ and $\hat \Sigma_{ij} = \COV(\tilde y_i, \tilde y_j)$.

\begin{lemma} \label{lemma:sandell}
  For any $\{S_{ij}\}_{i,j \in N}$, $\{P_{ij}\}_{i,j \in N}$ and $\{L_i\}_{i \in
  N}$ of compatible dimensions, the following matrix equation
  \begin{equation} \label{eq:generic-matrix-eq}
    \sum_{j \in N} \Big[
      S_{ij} F_j P_{ji} - S_{ij} L_j P_{ii}
    \Big] = 0,
    \quad \forall i \in N.
  \end{equation}
  for unknown $\{F_i\}_{i \in N}$ of compatible dimensions can be written in
  vectorized form as
  \begin{equation} \label{eq:vec}
    \Gamma F = \eta,
  \end{equation}
  where $F$, $\eta$, and $\Gamma$ are as defined in
  Theorem~\ref{thm:estimation}. Furthermore, define $S =
  [S_{ij}]_{i,j \in N}$ and $P = [P_{ij}]_{i,j \in N}$. 
  If $S > 0$, $P \ge 0$, and $P_{ii} > 0$, $i \in N$, then
  $\Gamma > 0$ and thus invertible. Then, Eq.~\eqref{eq:generic-matrix-eq} has
  a unique solution that is given by
  \begin{equation} \label{eq:generic-soln}
    F = \Gamma^{-1} \eta.
  \end{equation}
\end{lemma}
 The proof of Lemma~\ref{lemma:sandell} is presented in Appendix~\ref{sec:sandell-proof}.

\subsection{Proof of Theorem~\ref{thm:estimation}}

The key observation behind the proof is that Problem~\ref{prob:estimation} may be
viewed as a MTMSE filtering problem~\cite{Radner1962}, where agents observe
different information and want to minimize a common estimation cost.
For the ease of
notation, for a given agent~$i$, we let $(g_i, g_{-i})$ and $(\hat z_i, \hat
z_{-i})$ denote the strategy and estimates of all agents. 
Pick an agent $i \in N$, and fix the strategy $g_{-i}$ of all the other agents. Then the expected cost from the point 
of view of agent $i$ is given by
\[
\EXP\strut^{g_{-i}}[c(x, \hat z_i, \hat z_{-i}) | y_0, y_i ],
\]
where the superscript $g_{-i}$ in the expectation indicates that the cost
depends on the strategy of agents other than $i$. 

A necessary condition for optimality is that agent $i$ is 
playing a best response to the strategy of all other players, i.e., 
\begin{equation} \label{eq:sufficient-1}
  \frac{\partial}{\partial \hat z_i}
  \EXP\strut^{g_{-i}}[ c(x, \hat z_i, \hat z_{-i}) | y_0, y_i ] = 0, \quad
  \forall i \in N.
\end{equation}
It is shown in~\cite[Theorem
4]{Radner1962}, that when $c(x, \hat z)$ is convex,~\eqref{eq:sufficient-1} is
also a sufficient condition for optimality.

From the dominated convergence theorem, we can interchange the order of derivative
and expectation to get
\begin{align*}
  \hskip 1em & \hskip -1em
  \text{LHS of~\eqref{eq:sufficient-1}} =
  \EXP\strut^{g_{-i}}\bigg[ \frac{\partial}{\partial \hat z_i}
    c(x, \hat z_i, \hat z_{-i})
  \,\bigg|\, y_0, y_i \bigg]
  \\
  &=
  \EXP\strut^{g_{-i}}\bigg[ \frac{\partial}{\partial \hat z_i}
     \sum_{k \in N} \sum_{j \in N} (L_k x-\hat z_k)^\TRANS S_{kj} (L_jx-\hat z_j)
    \,\bigg|\, y_0, y_i \bigg] \\
  &= 2 \EXP\strut^{g_{-i}} \bigg[ \sum_{j \in N} S_{ij} (L_jx - \hat z_j)
    \,\bigg|\, y_0, y_i \bigg].
\end{align*}
Substituting the above in~\eqref{eq:sufficient-1}, we get that a necessary and
sufficient condition for a strategy $(g_i, g_{-i})$ to be team optimal~is
\begin{equation}
  \sum_{j \in N} \Big[
    S_{ij} \EXP\strut^{g_j}[ \hat z_j \,|\, y_0, y_i ]
  - S_{ij} L_j \EXP[ x \,|\, y_0, y_i ]
  \Big] = 0,
  \quad \forall i \in N.
\end{equation}
Note here that the superscript $g_{j}$ in $\EXP\strut^{g_j}[ \hat z_j \,|\, y_0, y_i ]$ highlights that the expectation 
depends on the choice of $g_j$. There is no such dependence 
in $\EXP[ x \,|\, y_0, y_i ]$.
Thus, the strategy $g$ given by~\eqref{eq:opt-common-static} is optimal if and
only if
\begin{multline} \label{eq:sufficient-3}
  \smash[d]{\sum_{j \in N} \bigg[}
    S_{ij} \EXP\bigl[ F_j(y_j - \hat y_j) + L_j \hat x_0
    \bigm| y_0, y_i \big]
  \\ - S_{ij} L_j \EXP\bigl[ x \bigm| y_0, y_i\bigr]\smash{\bigg]} = 0,
  \quad \forall i \in N,
\end{multline}
or equivalently
\begin{multline} \label{eq:sufficient-4}
  \smash[d]{\sum_{j \in N} \bigg[}
    S_{ij} F_j \EXP\bigl[ \tilde y_j | y_0, y_i \big]
    \\[-5pt]
   - S_{ij} L_j \EXP\bigl[ x - \hat x_0 \bigm| y_0, y_i\bigr]
 \smash{\bigg]} = 0.
  \quad \forall i \in N.
\end{multline}
Note that from Lemma~\ref{lem:P}, we have
\[
  \EXP[ x - \hat x_0 | y_0, y_i ]
  % = \hat x_i - \hat x_0
  = \hat \Theta_i \hat \Sigma_{ii}^{-1}  \tilde y_i.
\]
Substituting the above and the expression for ${\EXP[\tilde y_j |y_0,y_i]}$ 
from Lemma~\ref{lem:P} in~\eqref{eq:sufficient-4}, we get
that the strategy given by~\eqref{eq:opt-common-static} is optimal if and only
if, for all $i \in N$,
\begin{equation*}
  \sum_{j \in N}
     \Big[
    S_{ij} F_j \hat \Sigma_{ji} \hat \Sigma_{ii}^{-1}
- S_{ij} L_j \hat \Theta_i \hat \Sigma_{ii}^{-1} \Big] \tilde y_i =0.
\end{equation*}
Since the above should hold for all $\tilde y_i \in \reals^{d_y^i}$, the coefficient
of $\tilde y_i$ must be identically zero. Thus, the strategy given
by~\eqref{eq:opt-common-static} is optimal if and only if
\begin{equation} \label{eq:sufficient-final-y}
  \sum_{j \in N}
     \Big[
    S_{ij} F_j \hat \Sigma_{ji} \hat \Sigma_{ii}^{-1}
- S_{ij} L_j \hat \Theta_i \hat \Sigma_{ii}^{-1} \Big]  =0,
  \quad \forall i \in N.
\end{equation}

Furthermore, Lemma~\ref{lemma:sandell} implies that  when 
$\hat \Sigma_{ii} > 0$, then~\eqref{eq:sufficient-final-y} has a unique solution given by~\eqref{eq:F-static}.

Now for the minimum value of the estimation error, consider a single term of
the estimation error
\begin{align}
  &%\hskip 0.5em & \hskip -0.5em
  \EXP[ (L_i x- \hat z_i)^\TRANS S_{ij} (L_j x - \hat z_j) ]
  \notag  \\
  % &= \EXP[
  %   (L_i x - F_i(y_i - \hat y_i) - L_i \hat x_0)^\TRANS S_{ij}
  %   (L_j x - F_j(y_j - \hat y_j) - L_j \hat x_0)
  % ] \notag \\
  &\stackrel{(a)}= \EXP\big[
    (x - \hat x_0)^\TRANS L_i^\TRANS S_{ij} L_j (x - \hat x_0)
    \notag \\
    & \qquad
    - 2 (y_i - \hat y_i)^\TRANS F_i^\TRANS S_{ij} L_j (x - \hat
    x_0)
    \notag \\
    & \qquad +
     (y_i - \hat y_i)^\TRANS F_i^\TRANS S_{ij} F_j (y_j - \hat y_j)
    \big]
    \displaybreak[3]
    \notag \\
    &\stackrel{(b)}= \Tr(P_0 L_i^\TRANS S_{ij} L_j)
     - 2 \Tr(\hat \Theta_{i} F_i^\TRANS S_{ij} L_j)
   %  \notag \\
   %  & \quad
    + \Tr(\hat \Sigma_{ij}^\TRANS F_i^\TRANS S_{ij} F_j)
    \notag \\
    &\stackrel{(c)}= \Tr(P_0 L_i^\TRANS S_{ij} L_j)
    - 2 \Tr(F_i^\TRANS S_{ij} L_j \hat \Theta_{i})
   % \notag \\
   % & \quad
    + \Tr(F_i^\TRANS S_{ij} F_j \hat \Sigma_{ji}),
    \label{eq:single-term}
\end{align}
where $(a)$ follows from substituting~\eqref{eq:opt-common-static}, $(b)$
uses \hide{Lemma~\ref{lem:basic}, part~(2) and} Lemma~\ref{lem:P},
and $(c)$ uses the fact that for any matrices $\Tr(ABCD) =
\Tr(BCDA)$. Thus, the expected team estimation error is
\begin{align}
  J^* &= \sum_{i \in N} \sum_{j \in N}
  \EXP[ (L_i x - \hat z_i)^\TRANS S_{ij} (L_j x - \hat z_j) ] \notag \\
  &\stackrel{(d)}=
   \sum_{i \in N} \sum_{j \in N} \bigg[
     \Tr(P_0 L_i^\TRANS S_{ij} L_j)
     - 2 \Tr(F_i^\TRANS S_{ij} L_j \hat \Theta_{i})
    \notag \\
    & \hskip 6em
  + \Tr(F_i^\TRANS S_{ij} F_j \hat \Sigma_{ji}) \bigg]
  \notag \\
  &= \Tr(P_0 L^\TRANS S L) \notag \\
  &\quad - \sum_{i \in N} \Tr\Big( F_i^\TRANS \sum_{j \in N} \Big[
  2 S_{ij} L_j \hat \Theta_{i} -  S_{ij} F_j \hat \Sigma_{ji} \Big] \Big)
  \displaybreak[3]
  \notag \\
  &\stackrel{(e)}= \Tr(P_0 L^\TRANS S L)
   - \sum_{i \in N} \Tr\Big( F_i^\TRANS \sum_{j \in N}
   S_{ij} L_j \hat \Theta_{i}  \Big) \label{eq:single-term-2}
  %\notag \\
  %&\stackrel{(f)}= \Tr(P_0 L^\TRANS S L)
  %- \sum_{i \in N} \Tr( F_i^\TRANS S_i L \hat \Theta_{i})
\end{align}
where $(d)$ follows from~\eqref{eq:single-term}, and $(e)$ follows
from~\eqref{eq:sufficient-final-y}. %and $(f)$ uses the definition of $S_i$.
The result now follows from observing that
\begin{align*}
  \hskip 1em & \hskip -1em
   \sum_{i \in N} \Tr\Big( F_i^\TRANS \sum_{j \in N}
   S_{ij} L_j \hat \Theta_{i}  \Big)
   =
  \sum_{i \in N} \Tr(F_i^\TRANS S_i L \hat \Theta_{i}) \\
  &=
  \sum_{i \in N} \VVEC(F_i)^\TRANS \VVEC(S_i L \hat \Theta_{i})
  = F^\TRANS \eta = \eta^\TRANS \Gamma^{-1} \eta,
\end{align*}
where the first equality follows from $\Tr(A^\TRANS B) = \VVEC(A)^\TRANS
\VVEC(B)$.

\section{Proof of Lemma~\ref{lemma:sandell}} \label{sec:sandell-proof}

  By vectorizing both sides of~\eqref{eq:generic-matrix-eq} and using
  $\VVEC(ABC) = (C^\TRANS \otimes A) \times \VVEC(B)$, we get
  \[
    \sum_{j \in N} ( P_{ij} \otimes S_{ij} ) \VVEC(F_j)
    -
    \VVEC(S_{i \bullet} L P_{ii}) = 0,
    \quad \forall i \in N.
  \]
  Substituting $\Gamma_{ij} = P_{ij} \otimes S_{ij}$ and
  $\eta_i = \VVEC(S_{i \bullet} L P_{ii})$, we get~\eqref{eq:vec}.

  If $S > 0$, $P \ge 0$, and $P_{ii} > 0$, $i \in N$, then~\cite[Lemma 1]{Sandell1974} implies that
  $\Gamma > 0$ and thus invertible. Hence, Eq.~\eqref{eq:generic-matrix-eq}
  has a unique solution that is given by~\eqref{eq:generic-soln}.

\section{Proof of Theorem~\ref{thm:strategy-inf}} \label{sec:inf-proof}

$\bar \Sigma_{ii}$ is the variance of the innovation in the standard Kalman
filtering equation and by positive definiteness of $R_i$ is positive definite.
Lemma~\ref{lemma:sandell} implies that~\eqref{eq:matrix-eq-inf}
has a unique solution that is given by~\eqref{eq:F-inf}. To show the 
strategy~\eqref{eq:strategy-main-inf} is optimal, we proceed in two steps. 
We first identify a lower bound in optimal performance and then show 
that the proposed strategy achieves that lower bound.
\paragraph*{Step 1}
  From Theorem~\ref{thm:finite}, for any strategy $g$, we have that
  \[
  \frac{1}{T} J_T(g) \geq \frac{1}{T} \sum_{t=1}^{T} \big[
    \Tr(L^\TRANS S L P_0(t))
    -\eta(t)^\TRANS \Gamma(t) \eta(t)
  \big]
  \]
  Taking limits of both sides and using Lemma~\ref{lemma:covariance-inf}
  (which implies that $\lim_{t \to \infty} \eta(t) = \bar \eta$ and $ \lim_{t
  \to \infty} \Gamma(t) = \bar \Gamma$), we get
  \begin{equation} \label{eq:lower-bound}
    \limsup_{T \rightarrow \infty} \frac{1}{T} J_T(g)   \ge
  \Tr(L^\TRANS S L \bar P_0)
  -\bar \eta^\TRANS \bar \Gamma \bar \eta = J^*
  \end{equation}

\paragraph*{Step 2} Suppose $\hat z(t)$ is chosen according to
strategy~\eqref{eq:F-inf} and let $J(t)$ denote $\EXP[c(x(t), \hat z(t))]$.
Following~\eqref{eq:single-term} and~\eqref{eq:single-term-2} in the proof of
Theorem~\ref{thm:estimation}, we have that
\begin{align*}
 J(t) &= \Tr(L^\TRANS S LP_0(t) ) \notag \\
  &\quad - \sum_{i \in N} \Tr\bigg( \bar F_i^\TRANS \sum_{j \in N} \Big[
  2 S_{ij} L_j \hat \Theta_i(t)
  - S_{ij} \bar F_j \hat \Sigma_{ji}(t)
  \Big] \bigg).
\end{align*}
From Lemma~\ref{lemma:covariance-inf}, we have that
\begin{align*}
  \lim_{t \to \infty} J(t) &= \Tr(L^\TRANS S L\bar P_0 ) \notag \\
  &\quad - \sum_{i \in N} \Tr\bigg( \bar F_i^\TRANS \sum_{j \in N} \Big[
      2 S_{ij} L_j \bar \Theta_i
      - S_{ij} \bar F_j \bar \Sigma_{ji}
  \Big] \bigg).
  \\
  &= \Tr(L^\TRANS S L\bar P_0 )
  -\bar \eta^\TRANS \bar \Gamma \bar \eta
  = J^*.
\end{align*}
Thus, by Cesaro's mean theorem, we get \linebreak
\(
  \lim_{T \rightarrow \infty} \frac{1}{T} \sum_{t=1}^T J(t) = J^*.
\)
Hence, the strategy~\eqref{eq:F-inf} achieves the lower bound
of~\eqref{eq:lower-bound} and is therefore optimal.

\section{One-step delayed observation sharing} \label{app:one-step-delayed}

\subsection{Problem statement}
In this section, we use the result of Theorem~\ref{thm:finite} to show the 
relationship between MTMSE filtering and control in 
delayed observation sharing
model~\cite{Kurtaran1974,Sandell1974,Yoshikawa1975}. The notation used in this
section is self-contained and consistent with the standard notation used in
decentralized stochastic control.

Consider a decentralized control system with $n$ agents, indexed by the set
$N = \{1,\dots, n\}$. The system has a state $x(t) \in \reals^{d_x}$. The initial
state $x(1) \sim N(0,\Sigma_x)$ and the state evolves as follows:
\begin{align}
x(t+1) &= A(t) x(t) + B(t) u(t) + w(t), \label{eq:dynamic}
\end{align}
where $A$ and $B$ are matrices of appropriate dimensions. $u(t)
=~\VVEC(u_1(t), \cdots, u_n(t))$, where $u_i(t) \in \reals^{d_u^i}$ is the control
action chosen by agent $i$, and $\{ w(t)\}_{t \geq 1}$, $w(t) \in \reals^{d_x}$ is
an i.i.d.\@ process with $w(t) \sim \mathcal{N} (0,\Sigma_w)$. Each agent observes
a noisy version $y_i(t) \in \reals^{d^i_y}$ of the state given by
\begin{align}
y_i(t) = C_i(t) x(t) + v_i(t) \label{equ:observation}
\end{align}
 where $\{ v_i(t)\}_{t \geq 1}$, $v_i(t) \in \reals^{d^i_y}$, is an i.i.d.\@
 process with $v_i(t) \sim (0,\Sigma^i_v)$. This may be written in a vector
 form as
 \begin{equation}
   y(t) = C(t) x(t) + v(t) \label{equ:observation-total},
\end{equation}
   where $C = \ROWS(C_1,\dots, C_n)$, $v(t) = \VVEC(v_1(t), \dots, v_n(t))$,
   and $y(t) = \VVEC(y_1(t), \dots, y_n(t))$.

   \textbf{Assumption 1:} The primitive random variables
   $(x(1), \allowbreak \{ w(t)\}_{t \geq 1}, \allowbreak \{ v_1(t)\}_{t \geq 1}, \dots,\allowbreak
   \{v_n(t)\}_{t \geq 1})$ are independent.

 In addition to its local observation $y_i(t)$, each agent also receives the
 one-step delayed observations of all agents. Thus, the information available
 to agent $i$ is given by
 \begin{equation} \label{eq:information-structure}
 I_i(t) \coloneqq \left\lbrace y_i(t), y(1{:}t-1), u(1{:}t-1) \right\rbrace.
 \end{equation}
 Therefore, agent $i$ chooses
 the control action $u_i(t)$ as follows.
 \begin{equation}
   u_i(t) = g_{i,t} (I_i(t)),
 \end{equation}
 where $g_{i,t}$ is the control laws of agent $i$ at time $t$. The collection
 $g = (g_1, \dots, g_n)$, where $g_i = (g_{i,1}, \dots, g_{i,T})$ is called
 the control strategy of the system. The performance of any control strategy $g$
 is given by
 \begin{multline} \label{eq:cost}
 J(g) = \EXP^g \Big[ \sum_{t=1}^{T-1} \left[ x(t)^\TRANS  Q x(t)
 + u(t)^\TRANS R u(t) \right] \\ + x(T)^\TRANS  Q x(T) \Big],
\end{multline}
where $Q$ is symmetric positive semi-definite matrix, $R$ is symmetric
positive definite matrix, and
the expectation is with respect to the joint measure on the system
variables induced by the choice of $g$.
\begin{problem} \label{prob:one-step}
   Given the system dynamics and the noise statistics, choose a control strategy
   $g$ to minimize the total cost $J(g)$ given by~\eqref{eq:cost}.
 \end{problem}

Problem~\ref{prob:one-step} is a decentralized stochastic control problem. 
In such problems there is no separation of estimation and control
 (see, for example~\cite{Sandell1974}). We show that this lack
of separation is due to the fact that the MTMSE
filtering strategy depends on the weight matrix of the estimation cost.

\subsection{Equivalence to MTMSE filtering}

We start with a basic property of linear quadratic models.
Let $P(1{:}T)$ denote the solution to the following backward Riccati 
equation. $P(T) = Q$ and for $t \in \{ T-1, \dots, 1\}$,
\begin{align*}
  P(t) &= Q+A^\TRANS P(t+1) A \\
  &\quad - A^\TRANS
  P(t+1) B (R + B^\TRANS P(t+1) B)^{-1} B^\TRANS P(t+1) A.
\end{align*}
Define 
\begin{align*}
  S(t) &= R + B^\TRANS P(t+1) B, \\
  L(t) &= S(t)^{-1} (B^\TRANS P(t+1) A).
\end{align*}
Then, we have the following.

\begin{lemma} \label{lemma:equivalence-1}
  For any control strategy $g$, define 
  \begin{equation}
    J^\circ (g) = \sum_{t=1}^{T-1}\EXP[(u(t) + L(t) x(t))^\TRANS S(t) (u(t)+L(t) x(t))].
  \end{equation}
  Then, a strategy $g$ that minimizes $J^\circ (g)$ also 
  minimizes $J(g)$.
\end{lemma}

\begin{proof}
  Following~\cite[Chapter 8, Lemma 6.1]{Astrom1970}, we can 
  show that the total cost $J(g)$ can be written as 
  \begin{multline}
    J(g) = \sum_{t=1}^{T-1} \EXP \big[ w(t)^\TRANS P(t+1) w(t) +
    x(1)^\TRANS P(1) x(1) \big] \\
    + \sum_{t=1}^{T-1} \EXP \big[ (u(t) + L(t) x(t))^\TRANS S(t)
    (u(t) + L(t) x(t)) \Big].
  \end{multline}
  The third term is equal to $J^\circ (g)$ and the first two terms 
  do not depend on the control strategy $g$. Thus, $J(g)$ 
  and $J^\circ (g)$ have the same argmin. 
\end{proof}

Now, we split the state $x(t)$ into a deterministic part $\bar x(t)$
and a stochastic part $\tilde x(t)$ as follows.
\(
  \bar x(1) = 0, \quad \tilde x(1) = x(1),
\)
and
\begin{align*}
  \bar x(t+1) &= A \bar x(t) + B u(t),
  &
  \tilde x(t+1) &= A \tilde x(t) + w(t), \\
  \bar y(t) &= C \bar x(t),
  &
  \tilde y(t) &= C \tilde x(t) + v(t).
\end{align*}

Since the system is linear, we have
\[
  x(t) = \bar x(t) + \tilde x(t) \quad \text{and} \quad
  y(t) = \bar y(t) + \tilde y(t).
\]

Note that $\bar x(t)$ is a function of the past control actions, which are
known to all agents. Now, for any control strategy~$g$, define $\hat z_i(t) =
u_i(t) + L_i(t) \bar x(t)$. Then, the cost $J^\circ(g)$ may be written as
\begin{equation} \label{eq:one-step-cost}
  \sum_{t=1}^{T-1} 
  \EXP[ ( \hat z_i(t) + L(t)\tilde x(t))^\TRANS S(t) (\hat z_i(t) + L(t)\tilde x(t))
  ].
\end{equation}
The process $\{\tilde x(t) \}_{t \ge 1}$ is an uncontrolled linear stochastic
process and the cost~\eqref{eq:one-step-cost} is of of the same form as the
weighted mean-square cost that we have considered in this paper. 

Following~\cite{Ho1972}, we define
\(
  \tilde I_i(t) = \{ \tilde y_i(t), \tilde y(1{:}t-1) \}
\)
which may be considered as the control-free part of the information
structure.
\begin{lemma} \label{lemma:static-reduction}
  For any strategy~$g$ and any agent $i \in N$, $\tilde I_i(t)$ is equivalent
  to $I_i(t)$, i.e., they generate the same sigma algebra.
\end{lemma}
\begin{proof}
  The result follows from a similar argument as given in~\cite[Chapter 7,
  Section 3]{Kumar1986}.
\end{proof}

Since $\tilde I_i(t)$ is equivalent to $I_i(t)$, we may assume that $\hat
z_i(t)$ is chosen as a function of $\tilde I_i(t)$ instead of $I_i(t)$. Thus,
Problem~\ref{prob:one-step} is equivalent to the following MTMSE filtering problem. 
\begin{problem} \label{prob:static}
  Suppose $n$ agents observe the linear dynamical system $\{ \tilde x(t) \}_{t
  \ge 1}$ and share their observations over a one-step delayed sharing
  communication graph. Thus, the information available at agent~$i$ is
  \[
    \tilde I_i(t) = \{ \tilde y_i(t), \tilde y(1{:}t-1) \}.
  \]
  Agent $i$ chooses an estimate $\hat z_i(t)$ of $\tilde x(t)$ according to an
  estimation strategy $h_{i,t}$, i.e., 
  \[
    \hat z_i(t) = h_{i,t} (\tilde I_i(t))
  \]
  to minimize an estimation cost given by~\eqref{eq:one-step-cost}.
\end{problem} 
Problem~\ref{prob:static} is a MTMSE filtering problem and can
be solved using Theorem~\ref{thm:finite}. One can then take
the solution of Problem~\ref{prob:static} and translate it back to
Problem~\ref{prob:one-step} as follows.

\begin{theorem} \label{thm:static}
  Let $h^*$ be the optimal strategy for Problem~\ref{prob:static}, i.e.,
  \begin{multline}
    h^*_{i,t} (\tilde I_i(t)) = - L_i(t) \hat{\tilde{x}}(t) \\
    - F_i(t) \Big(\tilde y_i(t) - \EXP[\tilde y_i(t)|\tilde y(1{:}t-1)]\Big),
  \end{multline}
  where 
  \begin{align*}
    \hat{\tilde{x}}(t) &= \EXP[\tilde x(t)|\tilde y(1{:}t-1)],
    \\
    L(t) &= \ROWS(L_1(t), \dots, L_n(t)) ,
  \end{align*}
  and the gains $\{F_i(t)\}$ are computed as per
  Theorem~\ref{thm:finite}. Define strategy $g^*$ as follows:
  \begin{equation} \label{eq:gstar}
    g^*_{i,t}(I_i(t)) = h^*_{i,t}(\tilde I_i(t)) - L_i(t) \bar x(t),
  \end{equation}
  i.e., 
  \begin{multline} \label{eq:gstar-2}
    g^*_{i,t}(I_i(t)) = - L_i(t) \hat x (t) \\
    - F_i(t) \Big(y_i(t) - \EXP[y_i(t)|y(1{:}t-1), u(1{:}t-1)]\Big),
  \end{multline}
  where $\hat x(t) = \EXP[x(t)|I^\com(t)]= \bar x(t) + \EXP[\tilde x(t)|\tilde y(1{:}t-1)]$.	
  Then $g^*$ is the optimal strategy for Problem~\ref{prob:one-step}.
\end{theorem}
\begin{proof}
  The change of variables $\hat z_i(t) = u_i(t) + L_i(t) \bar x(t)$ implies
  that if $h^*$ is an optimal strategy for Problem~\ref{prob:static}, then
  $g^*$ given by~\eqref{eq:gstar} is optimal for Problem~\ref{prob:one-step}. 

  To establish~\eqref{eq:gstar-2}, we need to show that $\hat x(t) = \bar x(t)
  + \hat {\tilde x}(t)$. Define, $I^\com(t) = \{ y(1{:}t-1), u(1{:}t-1)\}$ and
  $\tilde I^\com(t) = \{ \tilde y(1{:}t-1)\}$. Then by
  Lemma~\ref{lemma:static-reduction} we have, $I^\com(t)$ is equivalent to
  $\tilde I^\com(t)$, i.e., they generate the same sigma algebra. The rest of
  the proof follows from the definition of $\hat x(t)$. We have
  \begin{align*}
    \hat x(t) &= \EXP[x(t)|\tilde I^\com (t)] \\
    & \stackrel{(a)}{=} \EXP[\bar x(t)|I^\com (t)]+\EXP[\tilde x(t)|\tilde I^\com (t)]\\
    &\stackrel{(b)}{=} \bar x(t) + \hat{\tilde{x}}(t),
  \end{align*}
  where $(a)$ follows from state splitting and $I^\com (t) = \tilde I^\com
  (t)$ and $(b)$ follows from the fact that $\bar x(t)$ is a deterministic
  function of $I^\com (t)$. 
\end{proof}

The main take away is as follows. By a simple change of variables we showed
that the one-step delayed observation sharing problem is equivalent to a
MTMSE filtering problem, where the weight matrix $S(t)$ of the
estimation cost depends on the backward Riccati equation for the cost
function. The MTMSE filtering strategy depends on the
weight matrix $S(t)$ and that is the reason why there is no separation between
estimation and control. Nonetheless, the optimal gains can be computed as
follows.
\begin{enumerate}
	\item Solve a Riccati equation to compute the weight 
	functions $S(1{:}T)$ and gains $L(1{:}T)$.
	\item Solve a Kalman filtering equation (which does not depend on $S(1{:}T)$) 
	to compute the covariances $\hat \Sigma(t)$ and $\hat \Theta(t)$ defined in 
	Theorem~\ref{thm:finite}. 
	\item Use $S(t)$, $L(t)$, $\hat \Sigma(t)$, and $\hat \Theta(t)$ to obtain the optimal 
	gains $F_i(t)$ by solving a system of matrix equations. 
	\item Using Theorem~\ref{thm:static} above, we can write the optimal 
	strategy $g^*_{i,t}$ in terms of $F_i(t)$ and $L_i(t)$.
\end{enumerate}

\section*{Acknowledgment}
The authors are grateful to Peter Caines, Roland Malhame, 
and Demosthenis Teneketzis for useful discussion and feedback.

\bibliographystyle{IEEEtran}
\bibliography{IEEEabrv,../../../../References/mybib}

\begin{IEEEbiography}[{\includegraphics[width=1in,height=1.25in,clip,keepaspectratio]{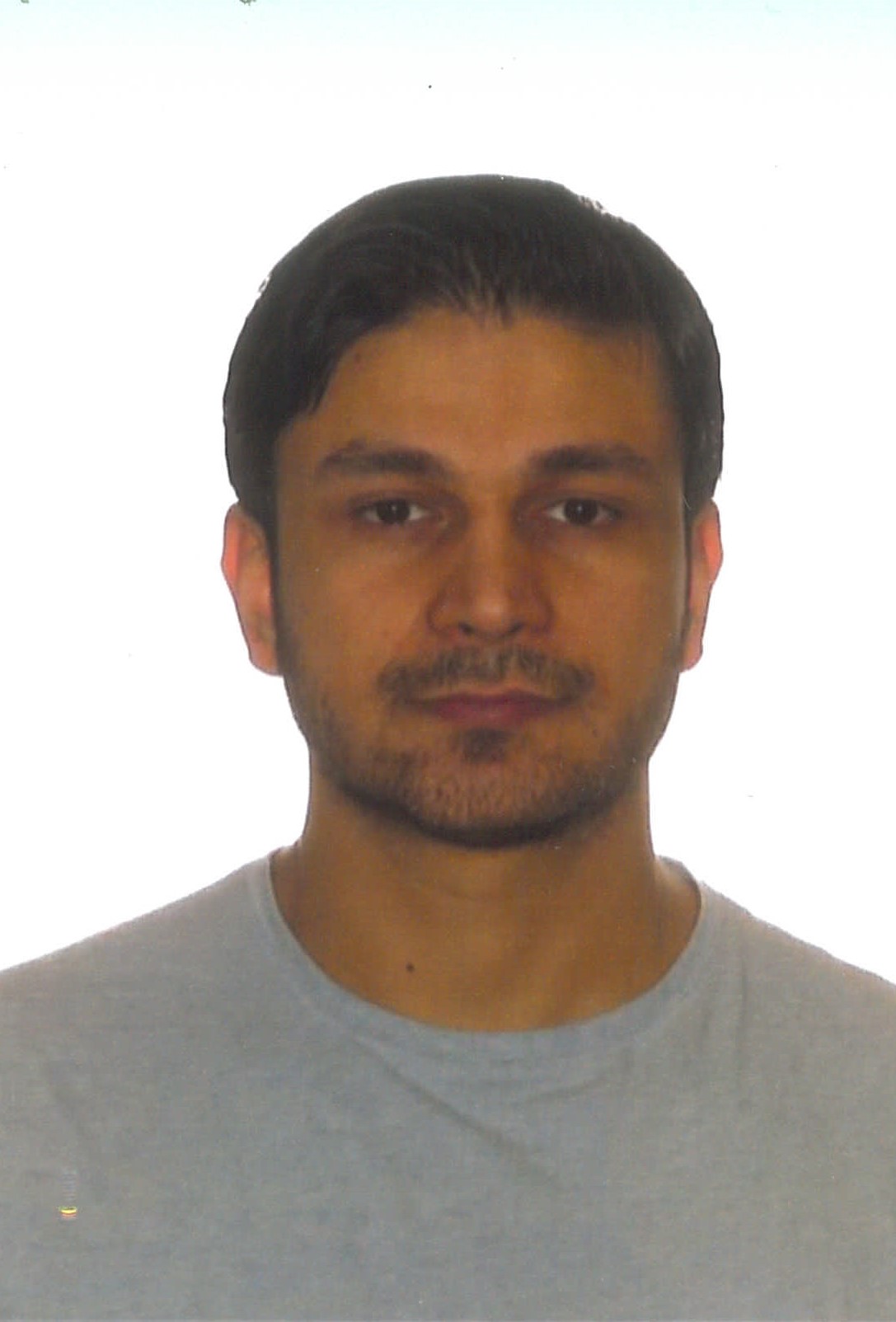}}]{Mohammad Afshari}
(S'12) received the B.S. and the M.S. degrees in Electrical Engineering from
the Isfahan University of Technology, Isfahan, Iran, in 2010 and 2012,
respectively. He is currently working towards the Ph.D. degree in Electrical
and Computer Engineering at McGill University, Montreal, Canada. His current area of research is 
decentralized stochastic control, team theory, and reinforcement learning.

Mr. Afshari is member of the McGill Center of Intelligent Machines (CIM) and member of the Research Group in Decision Analysis (GERAD).
\end{IEEEbiography}

\begin{IEEEbiography}[{\includegraphics[width=1in,height=1.25in,clip,keepaspectratio]{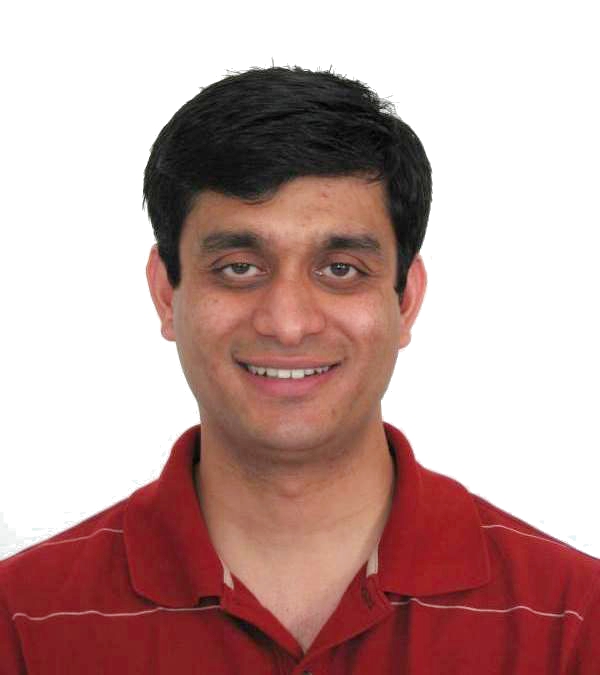}}]
  {Aditya Mahajan} (S’06-M’09-SM’14) received B.Tech degree from  the  Indian  Institute  of  Technology, Kanpur, India, in 2003, and M.S. and Ph.D. degrees from the University of Michigan, Ann Arbor, USA, in 2006 and 2008. From 2008 to 2010, he was a Postdoctoral Researcher at Yale University, New Haven, CT, USA. He has been with the department of  Electrical  and  Computer  Engineering,  McGill University, Montreal, Canada, since 2010 where he is currently Associate Professor. He serves as Associate Editor of Springer Mathematics of Control, Signal, and Systems. He was an Associate Editor of the IEEE Control Systems Society Conference Editorial Board from 2014 to 2017. He is the recipient of the 2015 George Axelby Outstanding Paper Award, 2014 CDC Best Student Paper Award (as supervisor), and the 2016 NecSys Best Student Paper Award (as supervisor). His principal research interests include decentralized stochastic control, team theory, multi-armed bandits, real-time communication, information theory, and reinforcement learning.
\end{IEEEbiography}

\end{document}